\documentclass[english]{article}
\usepackage[T1]{fontenc}
\usepackage[latin9]{inputenc}
\usepackage{geometry}
\usepackage{amsmath}
\usepackage{amssymb}
\usepackage{graphicx}
\usepackage{esint}
\usepackage[numbers]{natbib}

\makeatletter

\providecommand{\tabularnewline}{\\}

\usepackage[T1]{fontenc}

\usepackage{hyperref}
\usepackage{mciteplus}

\usepackage{color}
\usepackage[makeroom]{cancel}
\usepackage[normalem]{ulem}
\definecolor{pkcolor}{rgb}{0,0.1,0.7}
\definecolor{ascolor}{rgb}{1,0,1}
\definecolor{mscolor}{rgb}{1,0,0}
\definecolor{sscolor}{rgb}{0.0,0.4,0.8}

\newcommand\pkout{\marginpar{\color{pkcolor}$\clubsuit$}\bgroup\markoverwith{\color{pkcolor}{\rule[0.4ex]{2pt}{0.8pt}}}\ULon}
\newcommand\asout{\marginpar{\color{ascolor}$\heartsuit$}\bgroup\markoverwith{\color{ascolor}{\rule[0.4ex]{2pt}{0.8pt}}}\ULon}
\newcommand\msout{\marginpar{\color{mscolor}$\diamondsuit$}\bgroup\markoverwith{\color{mscolor}{\rule[0.4ex]{2pt}{0.8pt}}}\ULon}
\newcommand\ssout{\marginpar{\color{sscolor}$\spadesuit$}\bgroup\markoverwith{\color{sscolor}{\rule[0.4ex]{2pt}{0.8pt}}}\ULon}

\makeatother

\usepackage{babel}
\begin{document}

\title{Estimating nonlinear effects in forward dijet production in ultra-peripheral
heavy ion collisions at the LHC 
\date{}}

\author{P. Kotko$^1$, K. Kutak$^2$, S. Sapeta$^2$, A. M. Stasto$^1$, M. Strikman$^1$ \vspace{5pt}
\\\\
{\it \small $^1$  Department of Physics,   The Pennsylvania State University} \\{\it \small   University Park, PA 16802, United States}
\\\\
{\it \small $^2$ Institute of Nuclear Physics PAN}\\ {\small\it Radzikowskiego 152, 31-342 Krak\'ow, Poland}
}
\maketitle

\vspace{-30em}
\begin{flushright}
  IFJPAN-IV-2017-3
\end{flushright}
\vspace{25em}

\begin{abstract}
Using the framework that interpolates between the leading power limit
of the Color Glass Condensate and the High Energy (or $k_{T}$) factorization
we calculate the direct component of the forward dijet production
in ultra-peripheral $\mathrm{Pb}$-$\mathrm{Pb}$ collisions at CM
energy $5.1\,\mathrm{TeV}$ per nucleon pair. The formalism is applicable
when the average transverse momentum of the dijet system $P_{T}$
is much bigger than the saturation scale $Q_{s}$, $P_{T}\gg Q_{s}$,
while the imbalance of the dijet system can be arbitrary. The cross
section is uniquely sensitive to the Weizs\"acker-Williams (WW) unintegrated
gluon distribution, which is far less known from experimental data
than the most common dipole gluon distribution appearing in inclusive
small-$x$ processes. 
We have calculated cross
sections and nuclear modification ratios using WW gluon distribution obtained from the dipole
gluon density through the Gaussian approximation. The dipole gluon distribution
used to get WW was fitted to the inclusive HERA data with the nonlinear
extension of unified BFKL+DGLAP evolution equation. The saturation
effects are visible but rather weak for realistic $p_{T}$ cut on
the dijet system, reaching about $20\%$ with the cut as low as $6\,\mathrm{GeV}$.
We find that the LO collinear factorization with nuclear leading twist
shadowing predicts quite similar effects.
\end{abstract}

\section{Introduction}

High energy collisions of heavy ions provide unique opportunity to
investigate the quark-gluon plasma regime of QCD. In addition, they also offer 
a more direct insight into dense initial nucleus states.
 Relativistic nuclei
are in fact very strong sources of electromagnetic field, thus when
they collide at large impact parameters it is possible to study photon-nucleus
interactions. Such ultra-peripheral heavy ion collisions (UPC) can
be  investigated using the current LHC setup,  and there are plenty of unique
possibilities to  explore various aspects of nuclear physics~\citep{Baltz2008}
and small-$x$ regime~\citep{Strikman2007}. 
In particular, photon-nucleus interactions can shed some light
on certain aspects of the small-$x$ physics, in principle the saturation phenomenon \citep{Gribov1983}.

In the present work we will be focused on the dijet production in
UPC in the kinematic configuration which probes relatively small values of $x$.
The purpose of this analysis is two-fold. First, we give predictions
within a framework which incorporates a non-linear gluon saturation
phenomenon. For certain differential cross sections
we will also give predictions  using the collinear factorization with
the leading twist nuclear shadowing \citep{Frankfurt2005,Frankfurt2012}. Second,
the dijet configurations in $\gamma A$ collisions are sensitive to
 subtle QCD effects related to gluon distributions appearing
in saturation formalism; we describe this situation in some more details
later below. Thus, the second goal is to check if UPC can shed some
light on that subject. Previous study in the similar context for the
Electron Ion Collider was done in \citep{Zheng2014} using slightly
different formulation.

One of the aspects of jet production in hadron-hadron collisions (for recent
review see \cite{Sapeta2015}) is
that factorization theorem does not work for observables sensitive to
transverse momenta of partons, like azimuthal correlations. 
More precisely, the transverse momentum dependent (TMD) parton densities
are not universal \citep{Bomhof:2007xt}. This shows up as a process
dependence of Wilson line structure in the operators entering the
definitions of leading twist TMD parton distribution functions. On
the other hand, the non-operator approaches like the DDT formula \citep{Dokshitzer1978}
provide a way to access such observables in the leading logarithmic
approximation. The TMD factorizations are much stronger as they hold
up to leading power.
 In addition, the operator definitions of TMD gluon distributions 
are valid also for small values of $x$ and in general
 the non-universality still holds. This is in particular true in the saturation regime.
In the Color Glass Condensate (CGC) effective theory~\citep{Gelis:2010nm}
 which models the saturation phenomenon the non-universality 
it appears as a proliferation of color averages of many Wilson line
operators. 
 These correlators in principle parametrise a non-perturbative 
physics, playing a similar role to parton distributions.
In fact, 
 there is a connection to the TMD factorization:
for dijet production it has been shown that the leading power limit
of the CGC formulas corresponds to the TMD expressions, provided that
the CGC color averages are replaced by the hadronic matrix elements
\citep{Dominguez:2011wm}. Moreover, it appears that there are two
fundamental 
 objects describing the hadronic target.
On the TMD factorization side, there are two 
TMD gluon distributions, having the most elementary Wilson
line structure (see also \citep{Kharzeev:2003wz}). The first one
is
\begin{equation}
xG_{1}\left(x,k_{T}\right)=2\int\frac{d\xi^{-}d^{2}\xi_{T}}{(2\pi)^{3}P^{+}}\, e^{ixP^{+}\xi^{-}-i\vec{k}_{T}\cdot\vec{\xi}_{T}}\left\langle P\right|\text{Tr}\left\{ F^{+i}\left(\xi^{-},\xi_{T}\right)U^{\left[+\right]\dagger}F^{+i}\left(0\right)U^{\left[+\right]}\right\} \left|P\right\rangle \,,\label{eq:xG1def}
\end{equation}
where $F^{\mu\nu}\left(x\right)$ are gluon strength tensors in fundamental
representation and $U^{\left[+\right]}$ is the Wilson line joining
the space-time points $\xi$ and $0$ through the $+\infty$. The trace
is over the fundamental color space. The second TMD gluon distribution
is 
\begin{equation}
xG_{2}\left(x,k_{T}\right)=2\int\frac{d\xi^{-}d^{2}\xi_{T}}{(2\pi)^{3}P^{+}}\, e^{ixP^{+}\xi^{-}-i\vec{k}_{T}\cdot\vec{\xi}_{T}}\left\langle P\right|\text{Tr}\left\{ F^{+i}\left(\xi^{-},\xi_{T}\right)U^{\left[-\right]\dagger}F^{+i}\left(0\right)U^{\left[+\right]}\right\} \left|P\right\rangle \,,\label{eq:xG2def}
\end{equation}
where $U^{\left[-\right]}$ is the Wilson line that goes from $\xi$
to $0$ via $-\infty$. The appearance of the form of the gauge links
$U^{\left[\pm\right]}$ is related to the kind of collinear gluons
that are resummed: $U^{\left[+\right]}$ represents the final state
interactions, while $U^{\left[-\right]}$ the initial state interactions.
In case of $xG_{1}$ one can get rid of the Wilson lines by a suitable
choice of gauge (and boundary condition at $+\infty$ for transverse
gauge field components) giving a gluon number density interpretation
to that distribution. 
On the CGC side,
 those two gluon
distributions appear in two different contexts. The $xG_{1}$ appears
in the McLerran-Venugopalan (MV) model \citep{McLerran:1993ka} as
the gluon number density calculated in the semi-classical approximation,
thus it is often named the Weizs\"acker-Williams (WW) gluon distribution.
The $xG_{2}$ appears as a Fourier transform of the forward dipole
amplitude on the nucleus, thus it is named the `dipole' gluon distribution.
In general,  these are two independent quantities, but they can be related in a certain approximation,
 see
Eq.~(\ref{eq:gaussianaprox}) below.
 It is interesting, that most `simple' QCD processes
like inclusive DIS, semi-inclusive DIS or Drell-Yan all probe the
dipole gluon distribution. Thus, this gluon distribution is quite well constrained
from data. It is not the case for the WW gluon distribution which
has been only calculated from models. However, it has been shown in
\citep{Dominguez:2011wm} that the WW gluon distribution can be probed
when more complicated final states are considered, in particular dijets
in $\gamma A$ and $pA$ collisions. The dihadron correlations in $pA$ collisions were used to test these unintegrated gluon distributions within the saturation framework \citep{Stasto2011,Albacete2010a}. The advantage of $\gamma A$
collisions is that in the close to back-to-back dijet configuration
the WW gluon is probed directly. Note that the A-dependent Sudakov broadening of disbalance is present in the DGLAP formalism, see \cite{Guzey:2012jp}
 for the study of this effect in the massive neutral gauge boson production.  Study of this effect for dijet production  is beyond the scope of this paper.

 Although in the present work we focus mainly on the nonlinear saturation phenomena, the nuclear effects are not restricted
to saturation alone. In fact, before the onset of saturation the nuclear shadowing may be significant. It is a leading twist effect since the suppression of gluons is encoded in the collinear PDFs within the collinear factorization approach \citep{Frankfurt2012}.
Large gluon shadowing consistent with prediction of \citep{Frankfurt2012} was reported in the coherent $J/\Psi$ photoproduction of $\mathrm{Pb}$. 
For example the shadowing for $x=10^{-3}$, $Q^2 \sim 3\, \mathrm{GeV}^2 $ was found to be $\approx 0.6$, for the recent discussion and extensive list of references see \citep{Guzey2016} (for $J/\Psi$ production in CGC see \citep{Lappi2013} and \citep{Ducloue2015,Ducloue2016}).
 In the present work we estimate both effects,  the leading twist nuclear
shadowing and the saturation, for selected jet observables. The question whether
and how these should be combined remains open and is beyond the scope of 
this paper.  

We are considering the limit when typical transverse momenta are  much larger than $Q_s$.
In the limit  $k_t \lesssim Q_s$ production of leading particles/jets may be
suppressed very strongly. This effect may be responsible for the large
suppression of the  forward pion production  in $d-Au$
 collisions at BNL, for the summary and references see chapter 8 of Ref.~\citep{Frankfurt2012}.
How fast these effects die out at $k_t > Q_s$  still has to be investigated. 

 In the small-$x$ literature the gluon distributions with transverse momentum dependence are typically called
unintegrated gluon distributions (UGDs). This would suggest that integrating UGD one gets a collinear gluon PDF. 
This is in general not the case, especially for nucleus. For example one is unable to obtain the leading twist shadowing 
from the existing UGDs. Despite this fact, we shall follow the common small-$x$ terminology and continue using the term UGD.

From the above review it is clear that the study of dijets in UPC
collision is of great importance for better understanding of the
WW gluon distribution in the small $x$ regime. One of the goals of
this work is to investigate whether the present LHC kinematics can
give a restriction of that distribution. This is done, by a direct calculation
of the cross sections and nuclear modification factors for various
observables taking into account existing information on WW gluon distribution.

The work is organized as follows. In Section~\ref{sec:framework}
we describe the framework and ingredients necessary to compute the
process of interest within the saturation regime. Next, in Section~\ref{sec:Setup}
we give detailed description of the unintegrated gluon distribution
functions, kinematic cuts and actual implementation of the formalism.
The results of numerical simulations are given in Section~\ref{sec:Results}.
Finally, we give a brief summary in Section~\ref{sec:Summary}.

\section{Factorization formula for the dijet cross section in UPC}
\label{sec:framework}

In the standard approach to ultra peripheral collisions one factorizes
the cross section into the quasi-real photon flux $dN_{\gamma}/dx_{\gamma}$
and the photon-nucleus cross section $d\sigma_{\gamma A\rightarrow2\,\textrm{jet}+X}$
\citep{Baltz2008}
\begin{equation}
d\sigma_{AA\rightarrow2\,\textrm{jet}+X}^{\mathrm{UPC}}=\int dx_{\gamma}\,\frac{dN_{\gamma}}{dx_{\gamma}}\, d\sigma_{\gamma A\rightarrow2\,\textrm{jet}+X}\,,\label{eq:UPCmaster1}
\end{equation}
where $x_{\gamma}$ is the fraction of the heavy ion longitudinal
momentum carried by the photon. The photon flux is given by the Weizs\"acker-Williams
approximation
\begin{equation}
\frac{dN_{\gamma}}{dx_{\gamma}}=\frac{2Z^{2}\alpha}{\pi x_{\gamma}}\left[\zeta K_{0}\left(\zeta\right)K_{1}\left(\zeta\right)-\frac{\zeta^{2}}{2}\left(K_{1}^{2}\left(\zeta\right)-K_{0}^{2}\left(\zeta\right)\right)\right]\,,\label{eq:photonflux}
\end{equation}
with $Z$ being the charge of the ion, $\alpha=1/137$ is the electromagnetic
coupling constant and $K_{0}$, $K_{1}$ are the modified Bessel functions.
Their argument is $\zeta=x_{\gamma}R_{A}\sqrt{S}/\gamma$, where $S$
is the CM energy squared, $\gamma$ is the Lorentz factor and $R_{A}$
is the nucleus radius.

We are interested in the $\gamma A\rightarrow2\,\textrm{jet}+X$ process
under the following assumptions: i) The nucleus is probed
at sufficiently small longitudinal momentum fraction $x_{A}$ so that
the saturation formalism applies.  In reality, as we show later, $x_A$ can reach values of order $\sim10^{-3}$  (this would require $p_T$ of the jet down to $10 \; {\rm GeV}$) so that we probably venture outside the applicability domain of the saturation formalism.  ii) We focus on the kinematic region where 
 $x_{A}<x_{\gamma}$ which implies that we look for a forward
dijet configuration along the photon direction. This restriction enforces
the system to be probed in a domain where saturation effects are more
pronounced. iii) The average transverse momentum of the dijet system
$P_{T}=\left(p_{T1}+p_{T2}\right)/2$ is much bigger than the saturation
scale $Q_{s}$, $P_{T}\gg Q_{s}$, and sets the hard scale of the
process: $\mu\sim P_{T}$. iv) The transverse disbalance of the dijets
$k_{T}=\left|\vec{p}_{T1}+\vec{p}_{T2}\right|$ can be anything allowed
by the kinematics -- this implies that the non-leading power corrections
have to be taken into account. 

Let us now describe the approach satisfying the above requirements
(the justification will be given later in this section). It is a straightforward
analog (in fact much simpler) of the formalism of \citep{Kotko:2015ura}
for dijets in $pA$ collisions. The following factorization
formula for the $\gamma A\rightarrow2\,\textrm{jet}+X$ process will be used: 
\begin{equation}
d\sigma_{\gamma A\rightarrow2\,\textrm{jet}+X}=\sum_{\left\{ q,\overline{q}\right\} }\int\frac{dx_{A}}{x_{A}}\int d^{2}k_{T}\, x_{A}G_{1}\left(x_{A},k_{T}\right)\, d\sigma_{\gamma g^{*}\rightarrow q\overline{q}}\left(x_{A},k_{T}\right)\,,\label{eq:yAfactoriz}
\end{equation}
where $xG_{1}$ is the Weizs\"acker-Williams UGD function. The partonic
cross section $d\sigma_{\gamma g^{*}\rightarrow q\overline{q}}$ is
calculated using the LO amplitude for the process $\gamma g^{*}\rightarrow q\overline{q}$,
where $g^{*}$ denotes the off-shell gluon. In the high energy approximation
the momentum of the gluon has only one longitudinal component, parallel
to the parent hadron. That is, taking the momentum of the nucleus
to be $p_{A}$, the momentum of $g^{*}$ is $k_{A}^{\mu}=x_{A}p_{A}^{\mu}+k_{T}^{\mu}$,
where $p_{A}^{\mu}=\left(1,0,0,-1\right)\sqrt{S}/2$ and $k_{T}^{\mu}=\left(0,k_{T}^{1},k_{T}^{2},0\right)$.
The gluon spinor index is projected on the vector $p_{A}$. It can
be shown that such amplitude is gauge invariant \citep{Catani:1990eg}.
In practical calculations we used the helicity amplitudes calculated
 from the program described in \citep{Kotko2014a} extended to quarks.
The two-particle phase space is  included in $d\sigma_{\gamma g^{*}\rightarrow q\overline{q}}$
and is constructed with the exact momentum conservation and taking into
account the transverse momentum $k_T$ of the  gluon.
 Let us note that in Eq.~(\ref{eq:yAfactoriz}) there is no hard scale dependence
in $xG_1$. In fact, in the saturation formalism the evolution with the hard
scale is not typically present. This is primarily because the formalism is
normally used for small hard scales, of the order of the saturation scale. In
our case however, the hard scale is set by the jet transverse momenta and may be
large. Thus, it is important to include the resummation of the Sudakov logarithms in the calculation. The way we do it is described at the end of this section.

Let us now justify the present approach. We shall show that Eq.~(\ref{eq:yAfactoriz}) coincides with known results in two regimes: in the linear regime with large dijet imbalance $k_{T}\sim P_{T}\gg Q_{s}$ and in the saturation regime with small imbalance $P_{T}\gg k_{T}\sim Q_{s}$.

Let us start with the first regime. The formula (\ref{eq:yAfactoriz})
is superficially identical to the High Energy factorization (HEF)
formalism \citep{Catani:1990eg,Catani:1994sq,Collins1991} for heavy
quark pair production in inclusive DIS. Namely, in the latter the
same phase space and off-shell matrix element is used. 
Consider now the limit $k_{T}\sim P_{T}\gg Q_{s}$ adequate to the HEF regime of
applicability. It can be shown that, for large $k_T$,
both  $xG_{1}$ and $xG_{2}$ have the same asymptotics \citep{Kharzeev:2003wz} (this is in fact true 
for many other TMD gluon distributions one can define at small $x$, see \citep{Marquet2016}).
Therefore, in the linear HEF regime there is just one universal UGD (as long as there is no hard scale dependence).
  This justifies Eq.~(\ref{eq:yAfactoriz}) in the regime $k_{T}\sim P_{T}\gg Q_{s}$.

Let us now consider another region of interest, i.e. $P_{T}\gg k_{T}\sim Q_{s}$.
In that region the saturation effects cannot be neglected, but the
power corrections $\mathcal{O}\left(k_{T}/P_{T}\right)$ can. Therefore,
as proposed in \citep{Dominguez:2011wm} let us  first consider the CGC regime,
$k_{T}\sim P_{T}\sim Q_{s}$, and take the leading power limit of the corresponding formulas. In
CGC picture the situation for two-particle production gets more complicated
comparing to inclusive production due to a more complicated color flow
in the dense nuclear matter.  The relevant formula reads \citep{Dominguez:2011wm}
\begin{multline}
\frac{d\sigma_{\gamma A\rightarrow2\,\textrm{jets}}}{d^{3}p_{1}d^{3}p_{2}}=\sum_{\left\{ q,\overline{q}\right\} }N_{c}\,\alpha \,e_{q}^{2}\,\delta\left(x_{B}p_{B}^{-}-p_{1}^{-}-p_{2}^{-}\right)\int\frac{d^{2}x_{1T}}{\left(2\pi\right)^{2}}\frac{d^{2}x'_{1T}}{\left(2\pi\right)^{2}}\frac{d^{2}x_{2T}}{\left(2\pi\right)^{2}}\frac{d^{2}x'_{2T}}{\left(2\pi\right)^{2}}\,\\
e^{-i\vec{p}_{1T}\cdot\left(\vec{x}_{1T}-\vec{x}'_{1T}\right)}e^{-i\vec{p}_{2T}\cdot\left(\vec{x}_{2T}-\vec{x}'_{2T}\right)}\psi_{\alpha\beta}^{\lambda}\left(\vec{x}_{1T}-\vec{x}_{2T}\right)\psi_{\lambda}^{\alpha\beta*}\left(\vec{x}'_{1T}-\vec{x}'_{2T}\right)\\
\left\{ 1+S_{x_{g}}^{\left(4\right)}\left(\vec{x}_{1T},\vec{x}_{2T};\vec{x}'_{2T},\vec{x}'_{1T}\right)-S_{x_{g}}^{\left(2\right)}\left(\vec{x}_{1T},\vec{x}_{2T}\right)-S_{x_{g}}^{\left(2\right)}\left(\vec{x}'_{2T},\vec{x}'_{1T}\right)\right\} \,,\label{eq:CGCformula1}
\end{multline}
where $\psi_{\alpha\beta}^{\lambda}$ is the wave function of the
$q\overline{q}$ dipole originating in photon with transverse polarization
$\lambda=1,2$ (see e.g. \citep{Dominguez:2011wm} for the precise
form) and the $S^{\left(i\right)}$ terms describe the interaction
of the dipole with the color nucleus field. They are given by the
correlators of Wilson lines $U\left(\vec{x}_{T}\right)=\mathcal{P}\exp\left\{ ig\int_{-\infty}^{+\infty}dx^{-}A_{a}^{+}\left(x^{-},\vec{x}_{T}\right)t^{a}\right\} $
with $t^{a}$ being the color generator in fundamental representation:
\begin{gather}
S_{x_{g}}^{\left(2\right)}\left(\vec{x}_{1T},\vec{x}_{2T}\right)=\frac{1}{N_{c}}\,\left\langle \mathrm{Tr}\, U\left(\vec{x}_{1T}\right)U^{\dagger}\left(\vec{x}_{2T}\right)\right\rangle _{x_{g}}\,,\\
S_{x_{g}}^{\left(4\right)}\left(\vec{x}_{1T},\vec{x}_{2T};\vec{x}'_{2T},\vec{x}'_{1T}\right)=\frac{1}{N_{c}}\,\left\langle \mathrm{Tr}\, U\left(\vec{x}_{1T}\right)U^{\dagger}\left(\vec{x}'_{1T}\right)U\left(\vec{x}'_{2T}\right)U^{\dagger}\left(\vec{x}{}_{2T}\right)\right\rangle _{x_{g}}\,.
\end{gather}
The operation $\left\langle .\right\rangle _{x_{g}}$ represents the
averaging over the distributions of color sources with $x_{g}$ being
the smallest $x$ probed, see e.g. \citep{Gelis:2010nm} for details.
Taking the leading power limit one gets \citep{Dominguez:2011wm}
\begin{equation}
\frac{d\sigma_{\gamma A\rightarrow2\,\textrm{jets}}}{d^{2}k_{T}}=\sum_{\left\{ q,\overline{q}\right\} }\int\frac{dx_{A}}{x_{A}}\,\, x_{A}G_{1}\left(x_{A},k_{T}\right)d\sigma_{\gamma g\rightarrow q\overline{q}}\left(x_{A}\right)\,,\label{eq:LTCGC}
\end{equation}
where 
the fraction $x_g$ was identified with $x_A$ 
 and  the CGC average of the appearing bilocal operator $F^{+i}\left(\xi_{1}^{-},\xi_{1T}\right)U^{\left[+\right]\dagger}F^{+i}\left(\xi_{2}^{-},\xi_{2T}\right)U^{\left[+\right]}$
was identified with the hadronic matrix element $\left\langle .\right\rangle _{x_{g}}\longleftrightarrow\left\langle P\left|.\right|P\right\rangle /\left\langle P|P\right\rangle $
so that there appears the exact definition of the Weizs\"acker-Williams
UGD $xG_{1}$ as given in Eq.~(\ref{eq:xG1def}). Let us note that the hard cross
section in Eq.~(\ref{eq:LTCGC}) is calculated on-shell in accordance
with the leading twist limit. Comparing Eqs.~(\ref{eq:yAfactoriz}) and
(\ref{eq:LTCGC}), we see that the latter is recovered by neglecting
the power corrections in $d\sigma_{\gamma g^{*}\rightarrow q\overline{q}}$
in Eq.~(\ref{eq:yAfactoriz}), thus in the regime $P_{T}\gg k_{T}\sim Q_{s}$
they coincide.

To summarize, the factorization formula (\ref{eq:yAfactoriz}) for
the $P_{T}\gg Q_{s}$ regime coincides with HEF when $k_{T}\sim P_{T}$
and with the leading power limit of CGC when $k_{T}\sim Q_{s}$. For
$k_{T}$ between those limiting values it provides a smooth interpolation
given by the off-shell matrix element and exact kinematics.

Let us note, that if the process under consideration was inclusive single jet production, Eq.~(\ref{eq:yAfactoriz}) would change. In that case,
the dipole gluon distribution $xG_2$ would appear instead of $xG_1$ \citep{Mueller1990,Kovchegov:2001sc,Kovner2001,Gelis2003}.

As already mentioned, there is an issue related to Eq. (\ref{eq:yAfactoriz}) or (\ref{eq:LTCGC}). Namely for $P_T\gg k_T$ the hard scale evolution is important and there should be some sort of the Sudakov form factor resumming large logarithms of the
form $\log\left(k_{T}/P_{T}\right)$.  
In the correlation regime of Eq.~(\ref{eq:LTCGC}), 
 it is rather simple -- the Sudakov form factor multiplies the r.h.s. This was applied in \citep{Zheng2014}
in the context of di-hadron correlations
at EIC.  
On more general ground, in the saturation formalism a comprehensive
study was done in \citep{Mueller2013}.
 In our study, which, as discussed, goes beyond the leading
power, we shall follow a different path. As we describe in the next
section, the formalism of Eq.~(\ref{eq:yAfactoriz}) is particularly convenient
to implement in a simple Monte Carlo program which is capable of generating
 complete kinematics, with full final state four momenta, and with unit weight
when necessary. Knowing the weights of particular events in a sample
one can estimate the Sudakov effect by a suitable modification of
the weights applying the Sudakov probability. From that point of view
it is independent of the saturation and can be applied for any sample
of events. In what follows we shall refer to this approach as the
Sudakov resummation model. In short, the model takes a weight $w_{i}\left(k_{T},P_{T}\right)$
for an event $i$ (we suppress other weight variables for brevity)
and modifies it by a surviving probability $\mathcal{P}\left(k_{T},P_{T}\right)$
of the gluon with transverse momentum $k_{T}$ which initiated the
event. This probability is related to the Sudakov form factor. It
is assumed that the gluons that do not survive at the scale $k_{T}$
appear at the scale $P_{T}$, so that the model does not change the
total cross section. See \citep{vanHameren:2014ala} for details.

\section{The framework for the  unintegrated gluon distribution function}

\label{sec:Setup}

The complete direct components of the UPC cross section (\ref{eq:UPCmaster1})
 have been implemented in the \textsf{LxJet} Monte Carlo program \citep{Kotko_LxJet}
based on \textsf{foam} algorithm \citep{Jadach:2002kn}. It was previously
used to calculate some jet observables within HEF (and beyond) in
$pp$ and $pA$ collisions, including forward-central and forward-forward
dijets \citep{vanHameren:2014ala,vanHameren:2014lna}, three-jet production
\citep{VanHameren2013}, $Z_{0}$+jet production \citep{VanHameren2015},
UGD fits \citep{Kotko:2015ksa} and recently forward-forward dijets
\citep{VanHameren2016} using the approach of \citep{Kotko:2015ura}.
The Sudakov resummation model is an independent plugin and is applied
on the top of the generated and stored events.

The crucial ingredient of the formula (\ref{eq:yAfactoriz}) is the WW UGD. As
discussed in the introduction, there are no existing experimental  constraints
for this gluon. Thus, in the present work we shall follow \citep{VanHameren2016}
and use the WW  distribution obtained from the dipole UGD using the Gaussian
approximation 

\begin{equation}
\nabla_{k_{T}}^{2}G_{1}\left(x,k_{T}\right)=\frac{4\pi^{2}}{N_{c}S_{\bot}(x)}\,\int\frac{d^{2}q_{T}}{q_{T}^{2}}\,\frac{\alpha_{s}(k_T^2)}{\left(\vec{k}_{T}-\vec{q}_{T}\right)^{2}}\,
xG_{2}\left(x,q_{T}\right)G_{2}\left(x,\left|\vec{k}_{T}-\vec{q}_{T}\right|\right)\,,
\label{eq:gaussianaprox}
\end{equation}
where $S_{\perp}(x)$ is the effective transverse area of the target.
 The Balitsky-Kovchegov (BK) evolution equation~\citep{Balitsky1996,Kovchegov:1999yj} is typically used to evaluate the dipole gluon distribution in the presence of saturation. 
In order to include subleading effects that may be important at non-asymptotic $x$ 
we have used a more involved Kwiecinski-Martin-Stasto equation~\citep{Kwiecinski:1997ee}
with the nonlinear term \citep{Kutak:2003bd} (below we set $xG_{2}\left(x,k_{T}\right)\equiv\mathcal{F}\left(x,k_{T}^{2}\right)$
for brevity):
\begin{multline}
\mathcal{F}\left(x,k_{T}^{2}\right)=\mathcal{F}_{0}\left(x,k_{T}^{2}\right)\\
+\frac{\alpha_{s}(k_T^2)N_{c}}{\pi}\int_{x}^{1}\frac{dz}{z}\int_{k_{T\,0}^{2}}^{\infty}\frac{dq_{T}^{2}}{q_{T}^{2}}\left\{ \frac{q_{T}^{2}\mathcal{F}\left(\frac{x}{z},q_{T}^{2}\right)\theta\left(\frac{k_{T}^{2}}{z}-q_{T}^{2}\right)-k_{T}^{2}\mathcal{F}\left(\frac{x}{z},k_{T}^{2}\right)}{\left|q_{T}^{2}-k_{T}^{2}\right|}+\frac{k_{T}^{2}\mathcal{F}\left(\frac{x}{z},k_{T}^{2}\right)}{\sqrt{4q_{T}^{4}+k_{T}^{4}}}\right\} \\
+\frac{\alpha_{s}(k_T^2)}{2\pi k_{T}^{2}}\int_{x}^{1}dz\,\left\{ \left(P_{gg}\left(z\right)-\frac{2N_{c}}{z}\right)\int_{k_{T\,0}^{2}}^{k_{T}^{2}}dq_{T}^{2}\mathcal{F}\left(\frac{x}{z},q_{T}^{2}\right)+zP_{gq}\left(z\right)\Sigma\left(\frac{x}{z},k_{T}^{2}\right)\right\} \\
-d\frac{2\alpha_{s}^{2}(k_T^2)}{R^{2}}\left\{ \left[\int_{k_{T}^{2}}^{\infty}\frac{dq_{T}^{2}}{q_{T}^{2}}\mathcal{F}\left(x,q_{T}^{2}\right)\right]^{2}+\mathcal{F}\left(x,k_{T}^{2}\right)\int_{k_{T}^{2}}^{\infty}\frac{dq_{T}^{2}}{q_{T}^{2}}\,\ln\left(\frac{q_{T}^{2}}{k_{T}^{2}}\right)\mathcal{F}\left(x,q_{T}^{2}\right)\right\} \,,\label{eq:KMS}
\end{multline}
where $\Sigma\left(x,k_{T}\right)$ is the accompanying singlet sea
quark distribution and $R$ has the interpretation of the target radius
(more precisely, it appears from the integration of the impact parameter
dependent gluon distribution assuming the uniform distribution of
gluons). The parameter $d$, $0<d\leq 1$ is set to $d=1$ for proton and can be varied for nucleus to study theoretical uncertainty. This equation accounts for DGLAP corrections, kinematic constraint
along the BFKL ladder and running strong coupling constant. Due to
all these (formally) sub-leading corrections this equation has been
proven to be useful in modeling more exclusive final states. The actual
initial condition $\mathcal{F}_{0}$ has been fitted to the inclusive DIS HERA \citep{Kutak:2012rf}. In what follows we shall name this set of UGD KS (Kutak-Sapeta). Also, the parameter
$R$ had to be fitted giving $R\approx2.4\,\mathrm{GeV}^{-1}$. The
set for a nucleus (actually the UGD per nucleon) is obtained by changing
the proton $R$ parameter according to the simple Woods-Saxon prescription
$R_{A}=A^{1/3}R$ where $A$ is the mass number. The nonlinear term
in (\ref{eq:KMS}) is enhanced then by $dA^{1/3}$ resulting in much
stronger saturation effects than in proton case (for $d=1$). In \citep{VanHameren2016}
except  $d=1$ for nucleus, also the values $d=\left\{ 0.5,0.75\right\} $ have been used to study the dependence of the results on the strength of
the nonlinear term. In the present work, the set with  $d=0.5$, corresponding to weaker saturation will be used.
As evident from Eq.~\eqref{eq:KMS} the saturation  effects will become important whenever the nonlinear term will be of the same order as the linear term. Thus one can characterize the strength of the nonlinear effects by the parameter defined as ratio of these two terms which is  proportional to the average gluon density per unit area, that is a saturation scale.  We choose $d=0.5$ to ensure that the saturation scale in the case of scattering off $\mathrm{Pb}$  is about 3 times larger than in the case of scattering off the proton. This choice  is consistent with the ratio of average gluon density in $\mathrm{Pb}$  and proton for $x\sim 10^{-3} \div 10^{-4} $  and  with  account of the leading twist nuclear shadowing, cf. Fig.~100, of \citep{Frankfurt2012}. The saturation scale for Pb for this value of $x$ is about  $Q^2_{sat}\sim 2 \div 3\, \mathrm{GeV}^2$, to be compared with $Q^2_{sat} \le1\, \mathrm{GeV}^2$ for the proton case.

For the $\mathrm{Pb}$ nucleus
the Woods-Saxon formula with the correction resulting from the $d$
parameter gives $R_{\mathrm{Pb}}\approx20.1\,\mathrm{GeV}^{-1}$. 
Keep in mind however that this number is only loosely related to
the true nuclear radius (in the sense that it is a radius within the
interpretation of the model of Eq.~(\ref{eq:KMS})). 

\begin{figure}
\begin{centering}
$\!\!\!\!\!\!\!$\includegraphics[width=7.1cm]{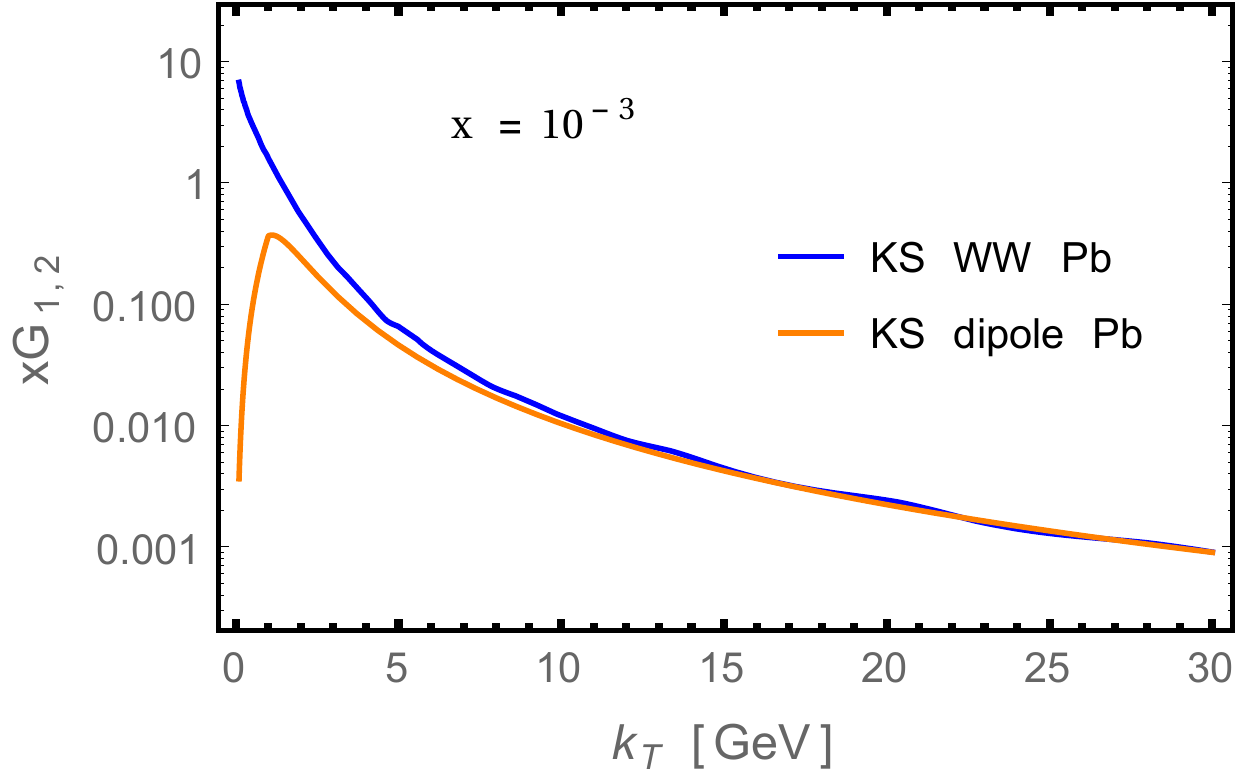}$\,\,\,\,$\includegraphics[width=7.1cm]{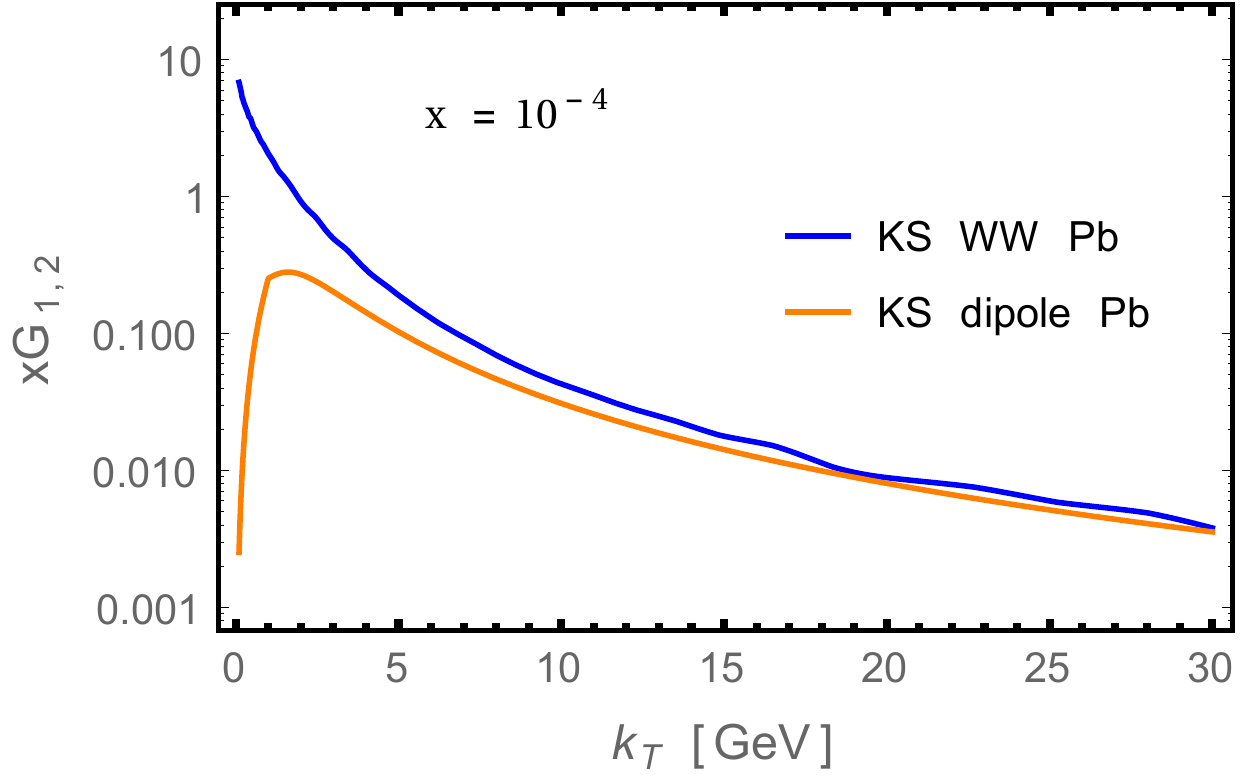}
\par\end{centering}
\vspace{1cm}

\begin{centering}
$\!\!\!\!\!\!\!$\includegraphics[width=7cm]{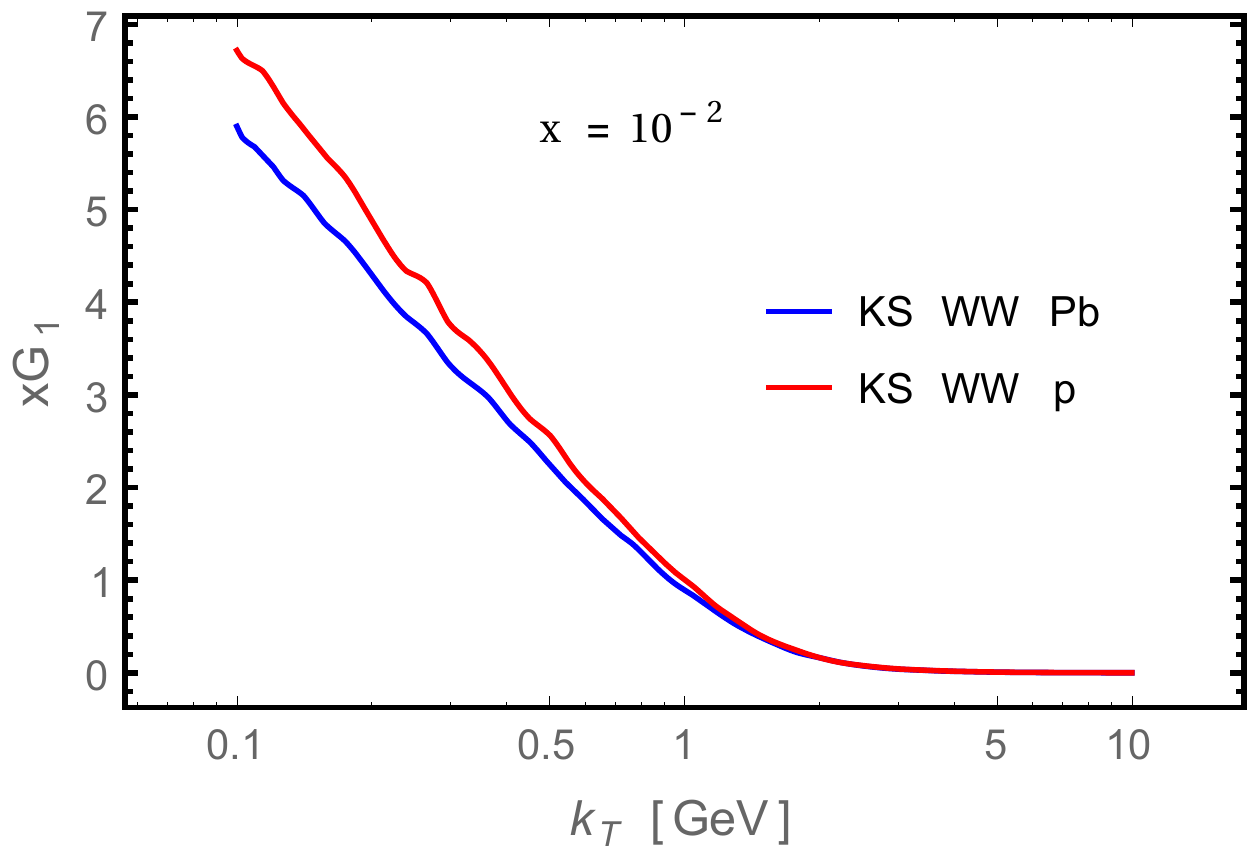}$\,\,\,\,$\includegraphics[width=7cm]{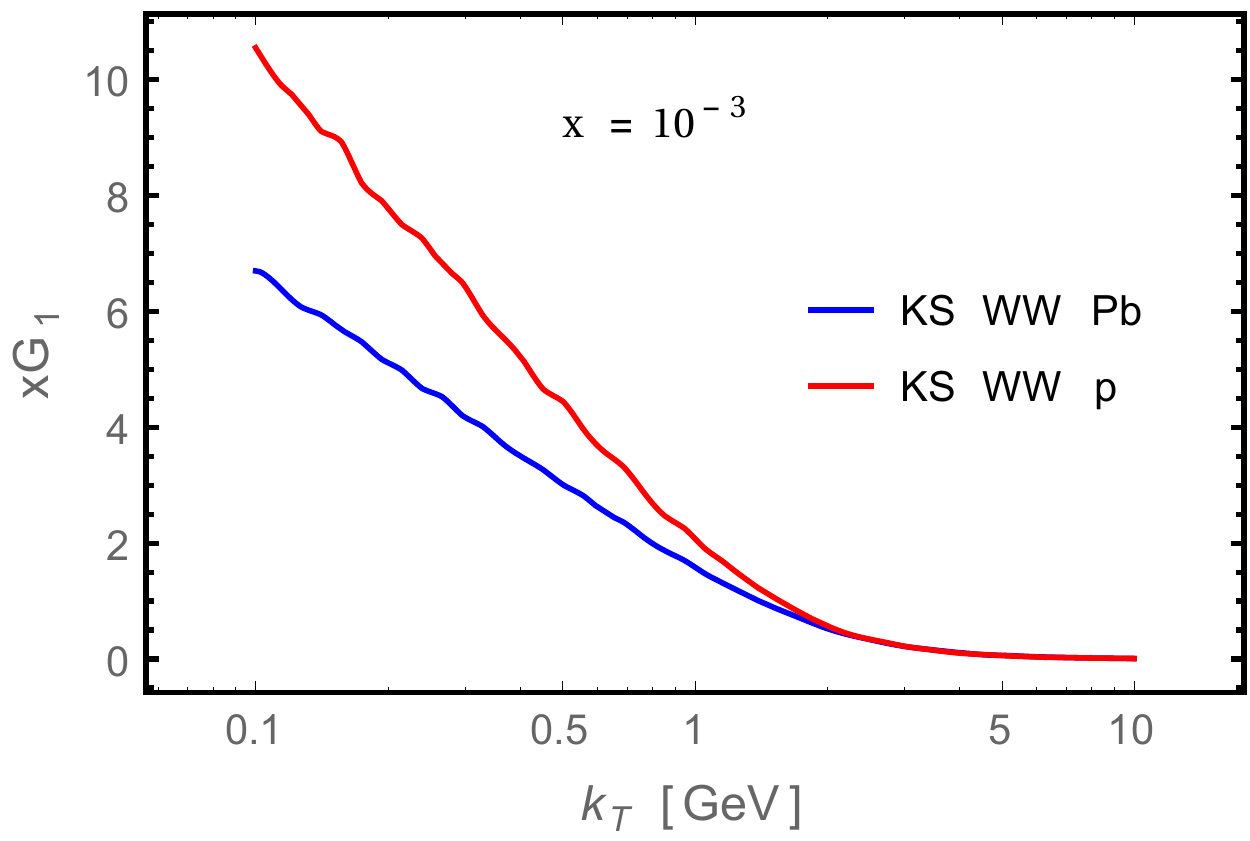}
\par\end{centering}
\vspace{1cm}

\begin{centering}
$\!\!\!\!\!\!\!$\includegraphics[width=7cm]{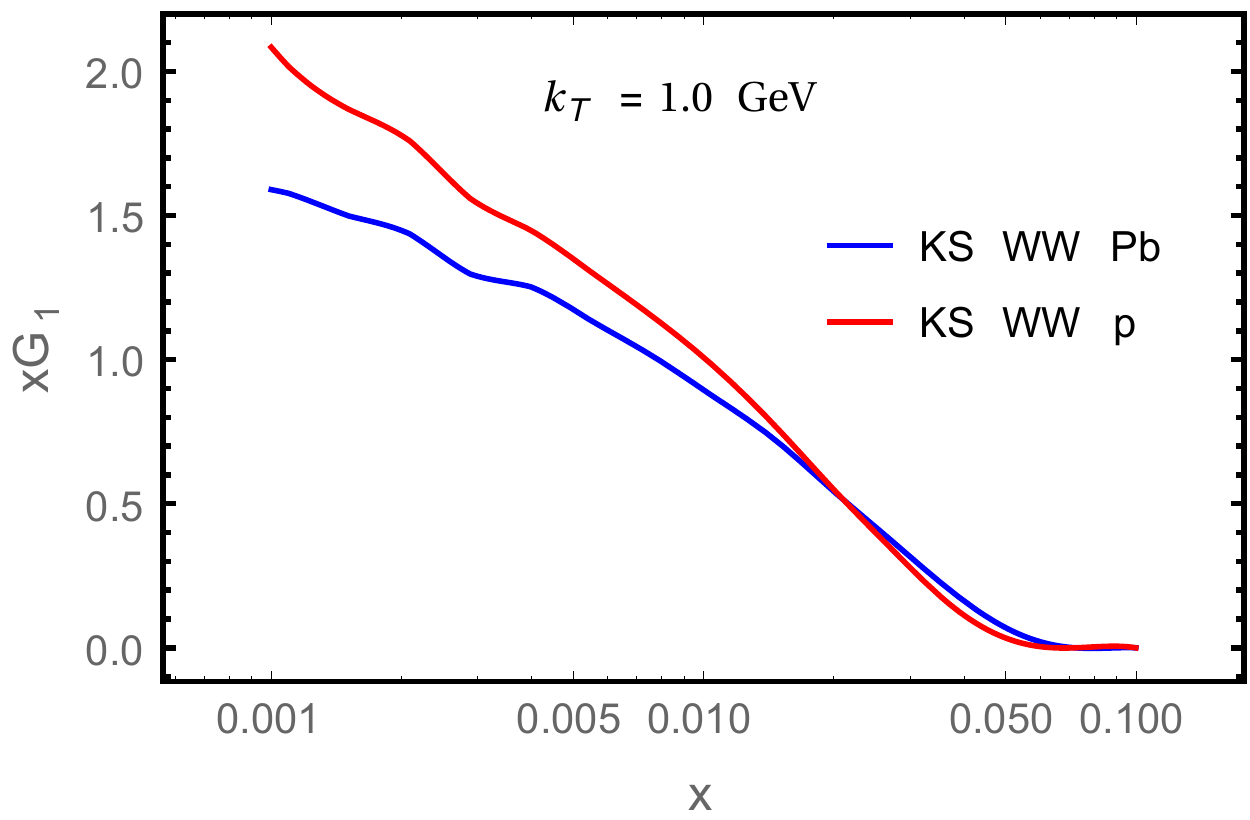}$\,\,\,\,$\includegraphics[width=7cm]{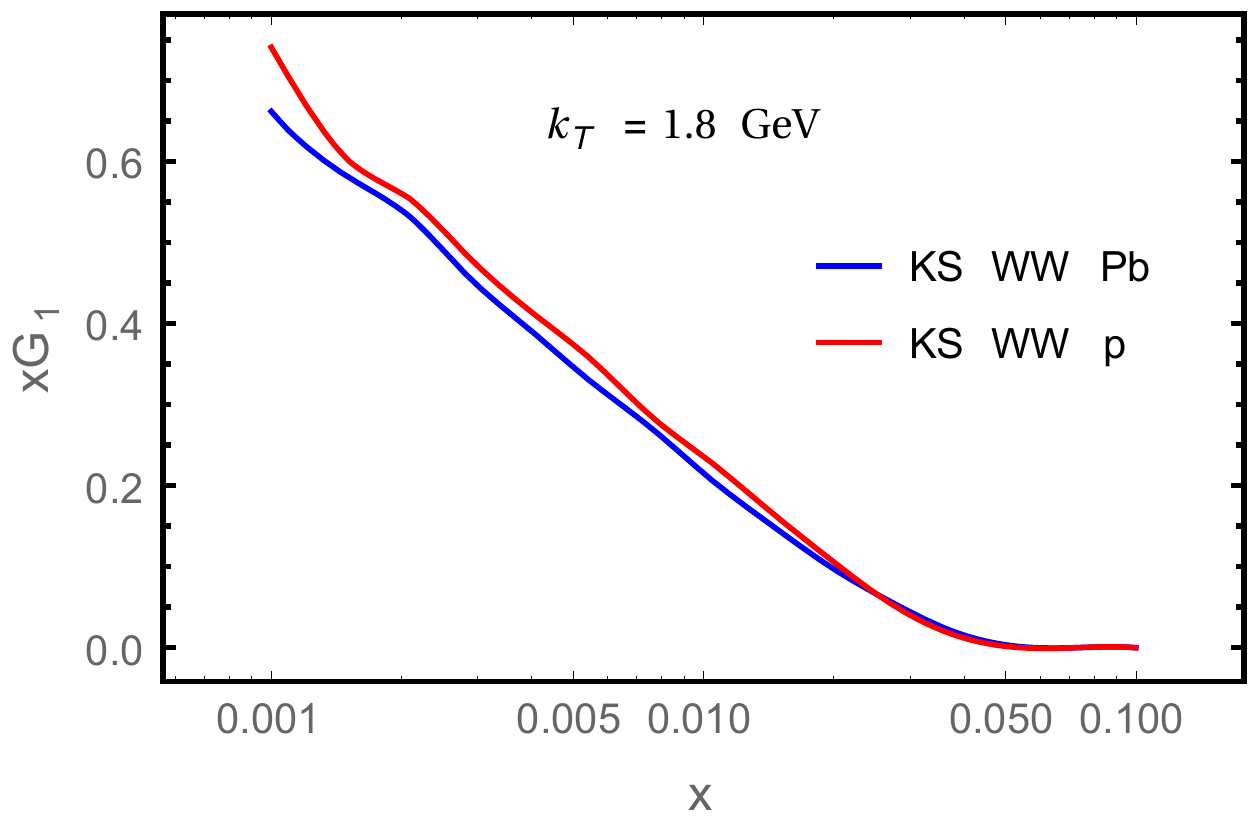}
\par\end{centering}

\caption{The Weizsacker-Williams (WW) unintegrated gluon distributions for proton and lead obtained from the KS dipole distributions \citep{Kutak:2012rf}. The top row compares the WW distribution for lead with the dipole distribution for two values of $x$. The middle row shows the WW distributions for proton and lead as a function of $k_T$ for two values of $x$. Finally, the bottom row shows  the WW distributions for proton and lead as a function of $x$ for two values of $k_T$.   \label{fig:WWUGD}}
\end{figure}

Finally, the $xG_{1}$ distributions for proton and lead were obtained from the
KS gluon $\mathcal{F}$ through Eq.~\eqref{eq:gaussianaprox}, using the method
presented in detail in \citep{VanHameren2016}.  We note, in particular, that in
our procedure of calculating the  WW KS distributions, $xG_{1}$, we used the
running coupling  and $x$-dependent transverse target area. The $x$ dependence of $S(x)$ was adjusted to ensure that the impact parameter dipole amplitude reaches unity in as expected in the black disk limit.
The resulting distributions are shown in Fig.~\ref{fig:WWUGD}. Let us note, that
 even for $x\sim 10^{-2}$ the nonlinear effects are still present in that model.
 This is one of the differences with respect to the leading twist 
 shadowing model and will be visible in the physical observables.

The kinematic setup is chosen as follows. We set the CM energy per nucleon to
$\sqrt{S}=5.1\,\mathrm{TeV}$ and require two jets in the rapidity window
$0<y_{1},y_{2}<5$ in the photon direction. 
The two-jet-requirement is assured by the jet algorithm of the
anti-$k_t$-type~\cite{Cacciari:2008gp}, which in the case of our two-particle
final state boils down to the condition $\sqrt{\Delta\phi^{2}+\Delta y^{2}}>r$,
where we choose the jet radius $r=0.5$. 
Note,
that although it is a LO calculation, the jet algorithm is necessary, because
the final states are in general not back-to-back, due to the transverse momentum
of the off-shell gluon. The minimal $p_{T}$ of the jets is dictated by the
longitudinal fractions $x$ we want to probe. Obviously, there is a competition
between the experimentally possible jet reconstruction and $x$ small enough to
see any saturation effects. In case of UPC, although the  CM energies of
$\gamma$-$A$ system are large, the photons cannot have too large longitudinal
momenta as above around $x\simeq0.03$ the flux becomes exponentially vanishing.
This considerably limits the $x$ fractions on the nucleus side, unless one goes
to really small $p_{T}$. In our study we have considered $p_{T}$ cuts of $25,$
$10$, and $6$ $\mathrm{GeV}$. As discussed in the next section with this setup
one can probe $x$ down to $10^{-3}$ at the current energy. The setup is summarized
in Table~\ref{tab:cuts}.

\begin{table}
\begin{centering}
\begin{tabular}{c|c}
CM energy & $\sqrt{S}=5.1\,\mathrm{TeV}$\tabularnewline
\hline 
rapidity & $0<y_{1},y_{2}<5$\tabularnewline
\hline 
transverse momentum & $p_{T1},p_{T2}>p_{T0}$,$\,\,$$\,$$p_{T0}=25,\,10,\,6\,\mathrm{GeV}$\tabularnewline
\end{tabular}
\par\end{centering}

\caption{The kinematic cuts used in the numerical calculations of the dijet
cross section in the ultra peripheral $\mathrm{Pb}-\mathrm{Pb}$ collisions.
\label{tab:cuts}}

\end{table}

\section{Numerical results}

\label{sec:Results}

\begin{figure}
\begin{centering}
\includegraphics[width=7cm]{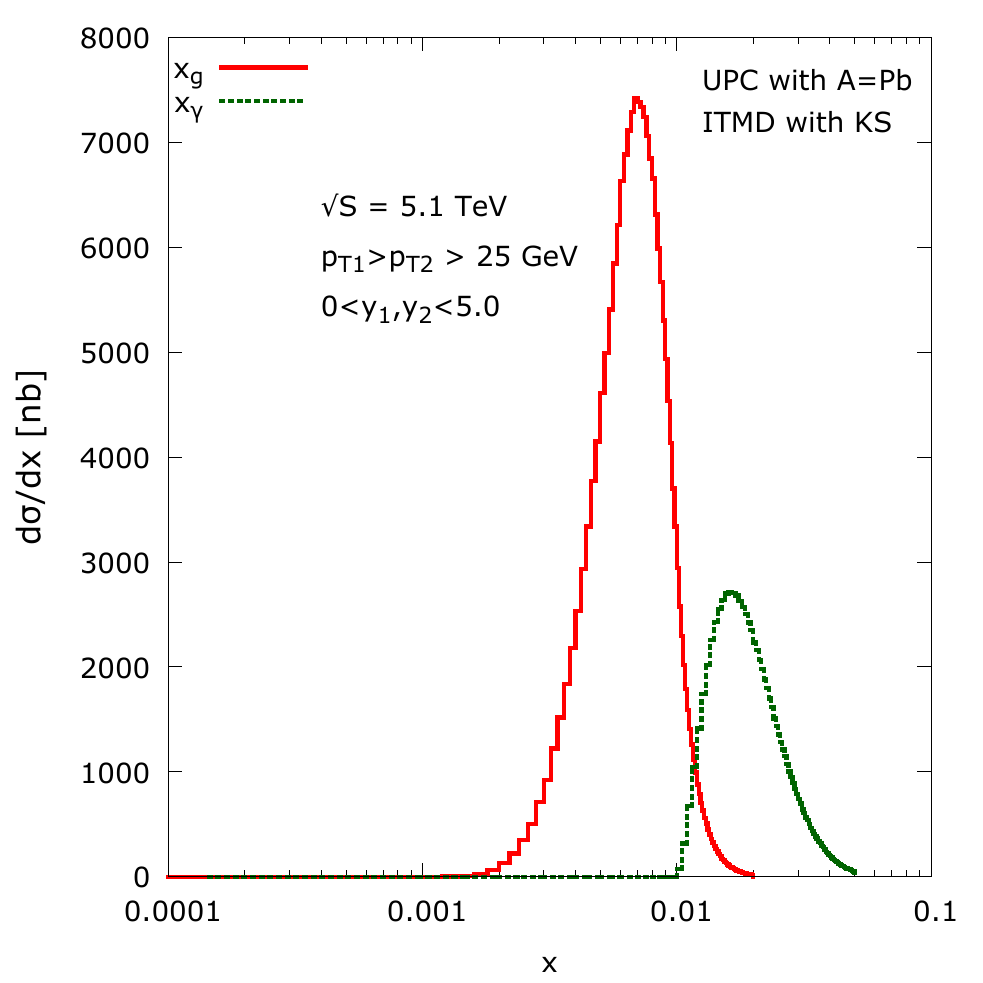}$\,\,$\includegraphics[width=7cm]{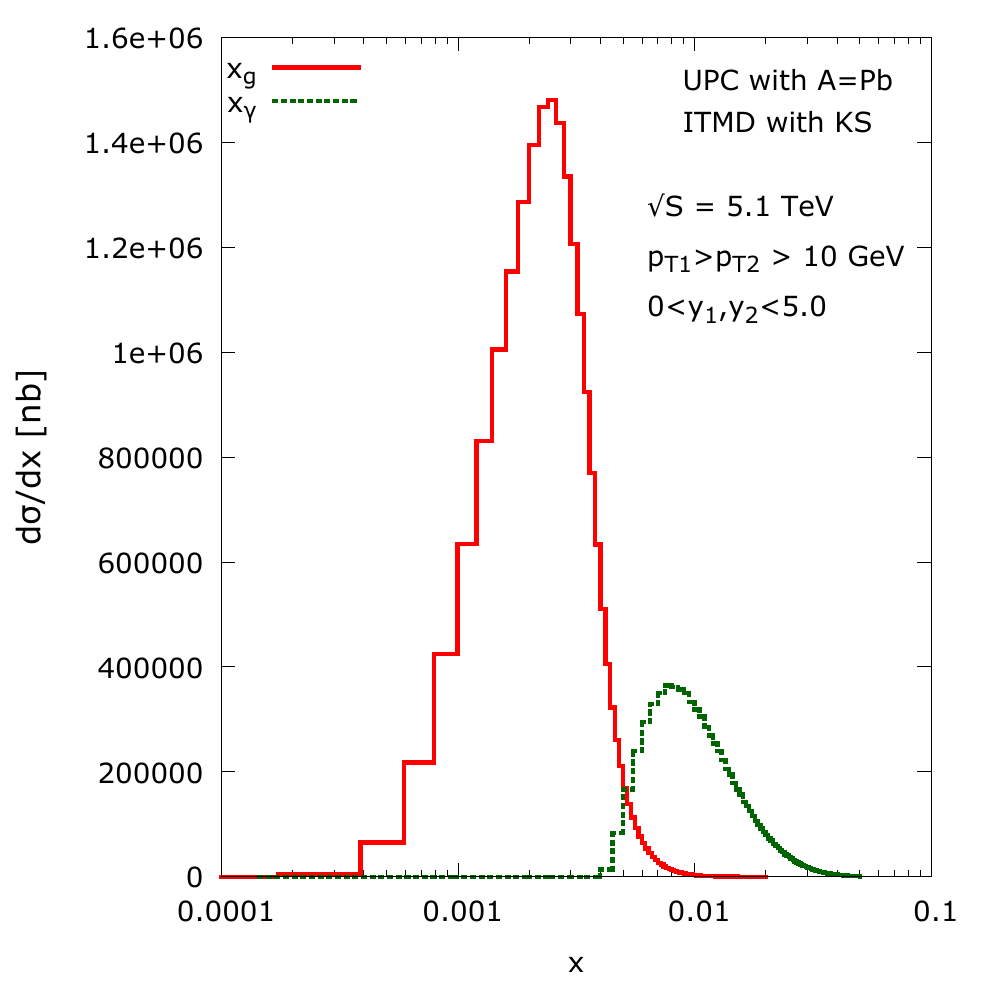}
\par\end{centering}

\begin{centering}
\includegraphics[width=7cm]{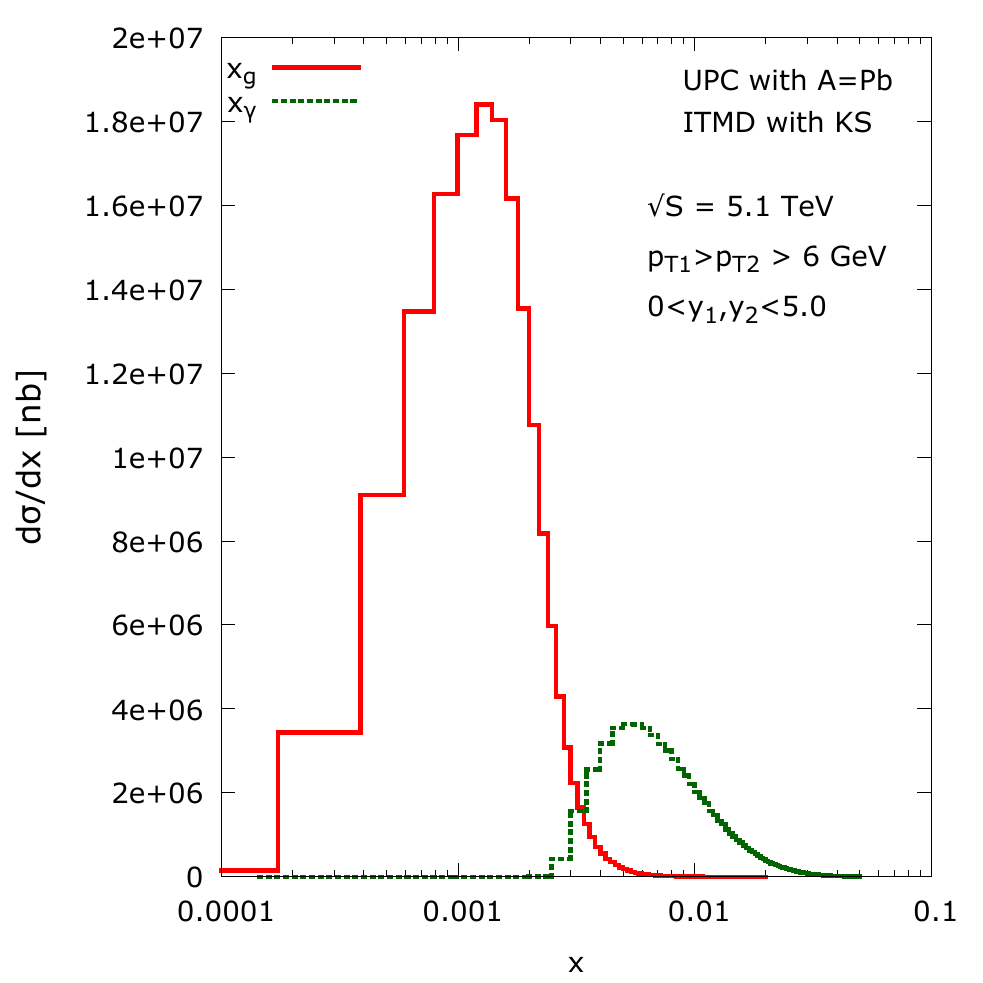}$\,\,$\includegraphics[width=7cm]{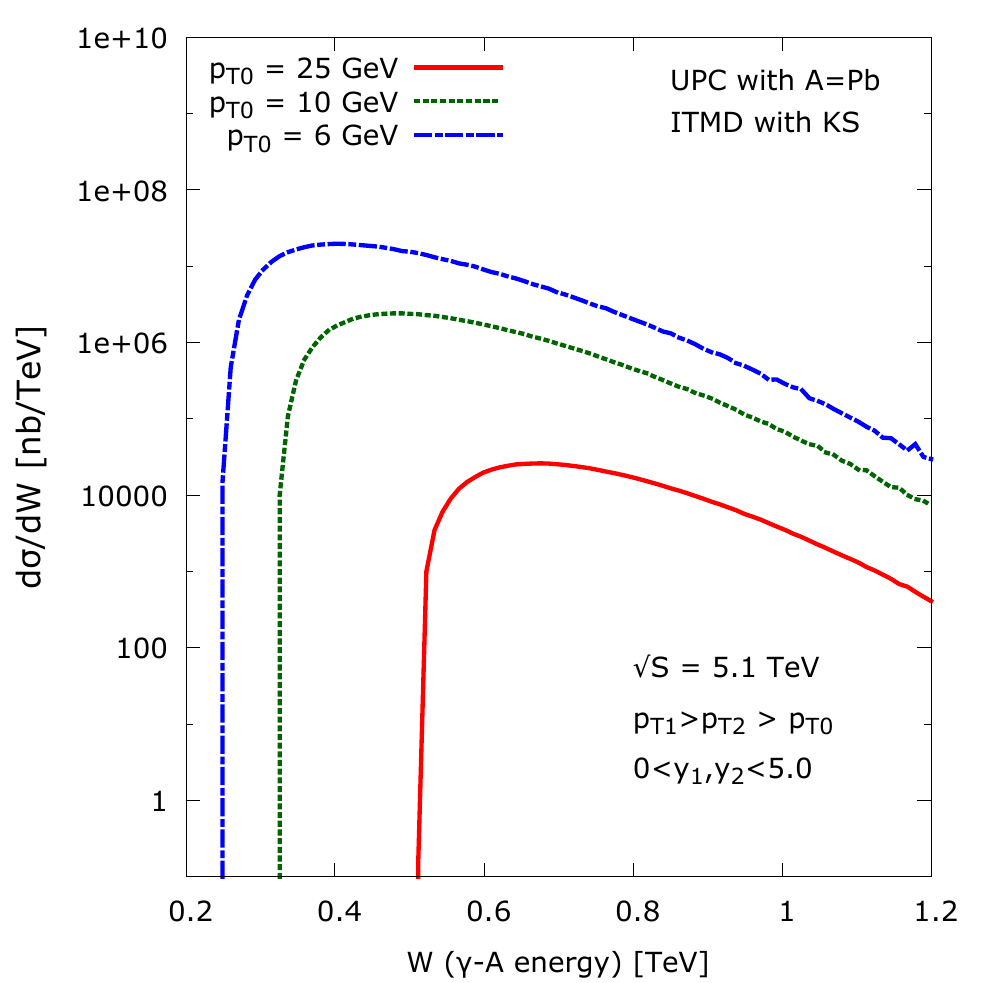}
\par\end{centering}

\caption{The longitudinal fractions probed within the kinematic cuts for different
$p_{T}$ cuts: top left $p_{T0}=25\,\mathrm{GeV}$, top right $p_{T0}=10\,\mathrm{GeV}$,
bottom left $p_{T0}=6\,\mathrm{GeV}$. In bottom right plot we show
the distribution of collision energy of $\gamma A$ system. \label{fig:xsec-xfrac}}

\end{figure}

We start by determining the longitudinal fractions $x$ of the photon and the gluon that can be
effectively probed within our cuts. In Fig.~\ref{fig:xsec-xfrac}
we show differential cross sections in the longitudinal fractions
for various $p_{T}$ cuts. We see, that for $p_{T0}=25\,\mathrm{GeV}$
the gluon longitudinal momentum fraction  $x_A$  is probed only slightly below $10^{-2}$, while  for $p_{T0}=10\,\mathrm{GeV}$
the process probes  $x_A$ easily around $10^{-3}$. With the smallest cut tested
$p_{T0}=6\,\mathrm{GeV}$ one can go below $10^{-3}$. 
We also show
the distribution of the $\gamma A$ CM energy, which reaches $1.2\,\mathrm{TeV}$. 
All these distributions are shown without the Sudakov effect, as its impact on these spectra is very weak.

\begin{figure}
\begin{centering}
\includegraphics[width=7cm]{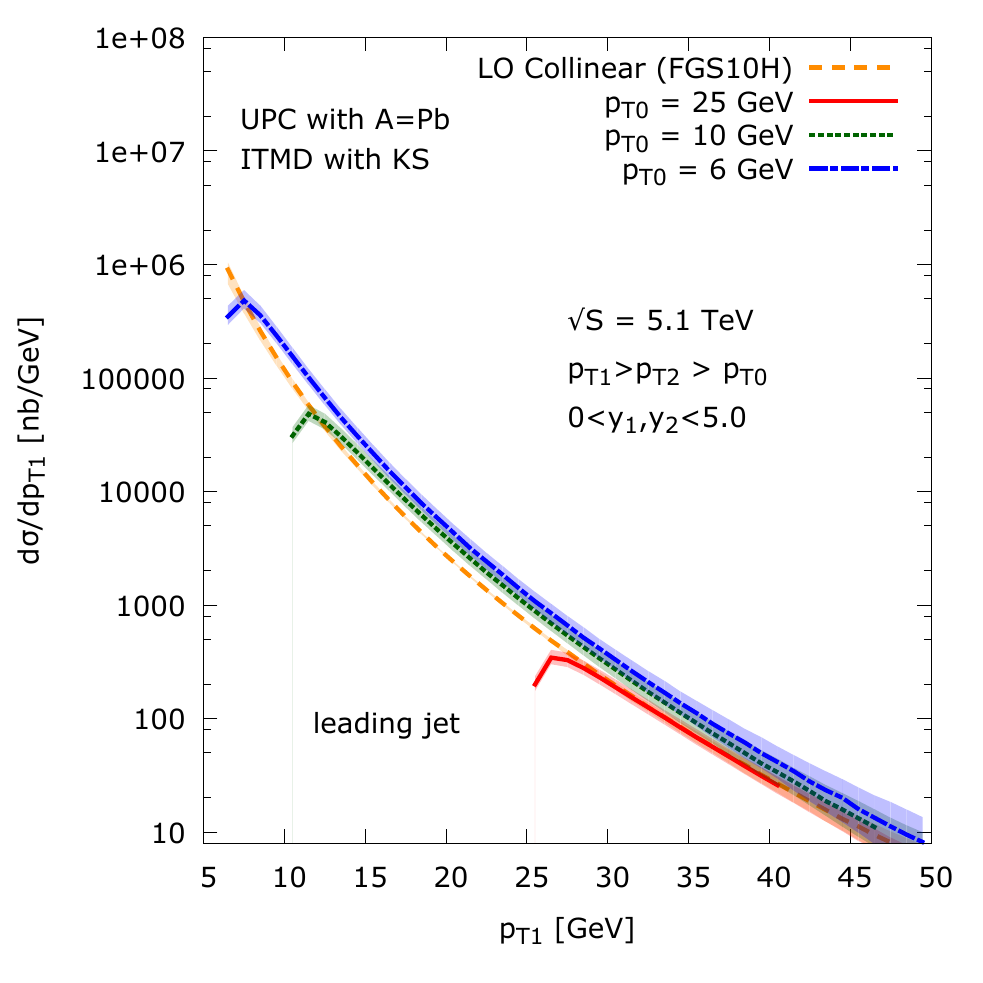}$\,\,$\includegraphics[width=7cm]{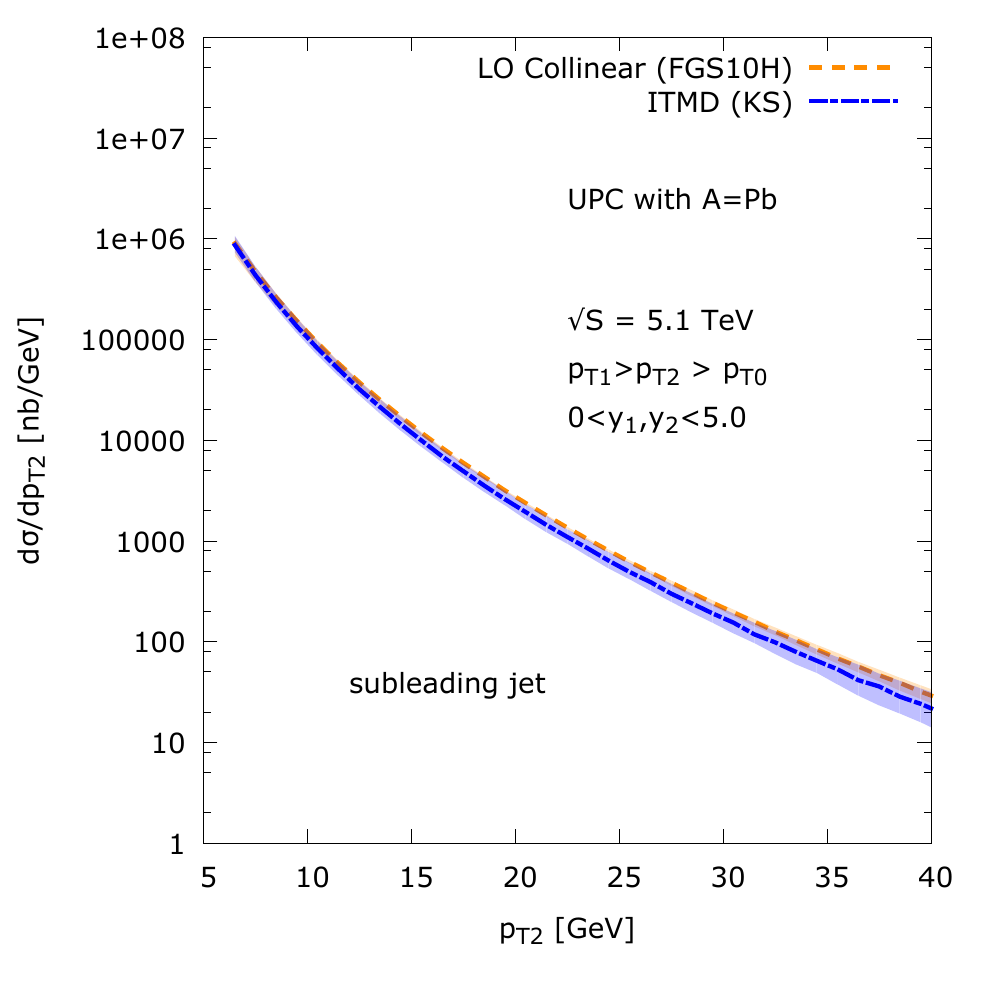}
\par\end{centering}

\begin{centering}
\includegraphics[width=7cm]{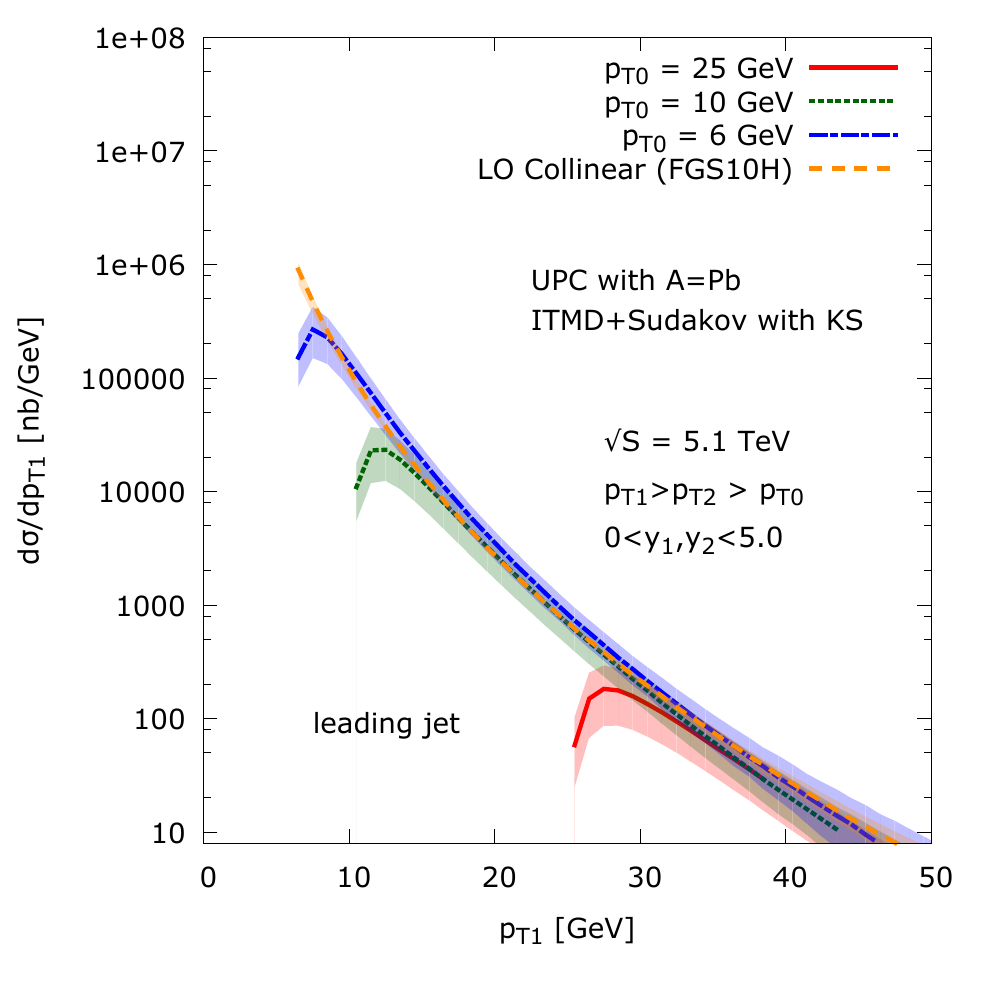}$\,\,$\includegraphics[width=7cm]{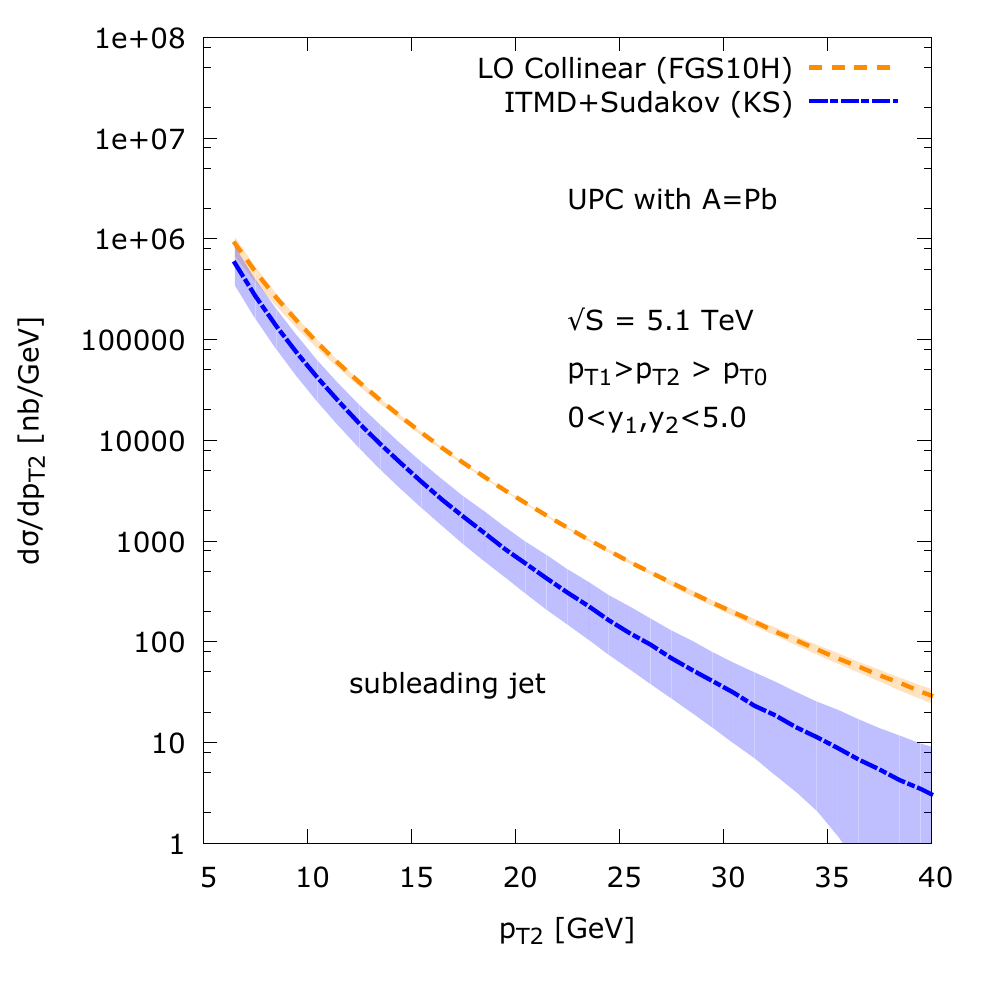}
\par\end{centering}

\caption{Transverse momentum spectra for leading (left column) and sub-leading
(right column) jets. The bottom row shows the effect of the Sudakov
resummation model. For comparison we show the results from the LO
collinear factorization using nuclear PDFs with the leading twist shadowing 
 (no additional effects are included in this LO calculation thus the spectra for leading and subleading jet are identical). 
\label{fig:xsec-pT}}
\end{figure}

In Fig.~\ref{fig:xsec-pT} we present the differential cross sections
in the jet $p_{T}$. In the present formalism the jets in general
do not have equal $p_{T}$ thus we order them, $p_{T1}>p_{T2}>p_{T0}$,
and show separate plots for leading and sub-leading jet spectra. For
comparison we also calculate the same observable from the LO collinear
factorization with nuclear PDFs implementing the leading twist nuclear
shadowing \citep{Frankfurt2005}. In the LO collinear factorization both
jets have equal $p_{T}$. Interestingly their spectra are very
close to the subleading jet spectrum of the present approach. The
error bands are constructed by varying the hard scale by the factor
of two with respect to the central value. The two bottom plots in
Fig.~\ref{fig:xsec-pT} show the effect of the Sudakov resummation
model. It has a significant effect on the subleading jet spectrum making
its slope bigger. The error bands are bigger with resummation because
the appearing hard scale can be varied as well and the results are sensitive to that scale. 

\begin{figure}
\begin{centering}
\includegraphics[width=8cm,height=6cm]{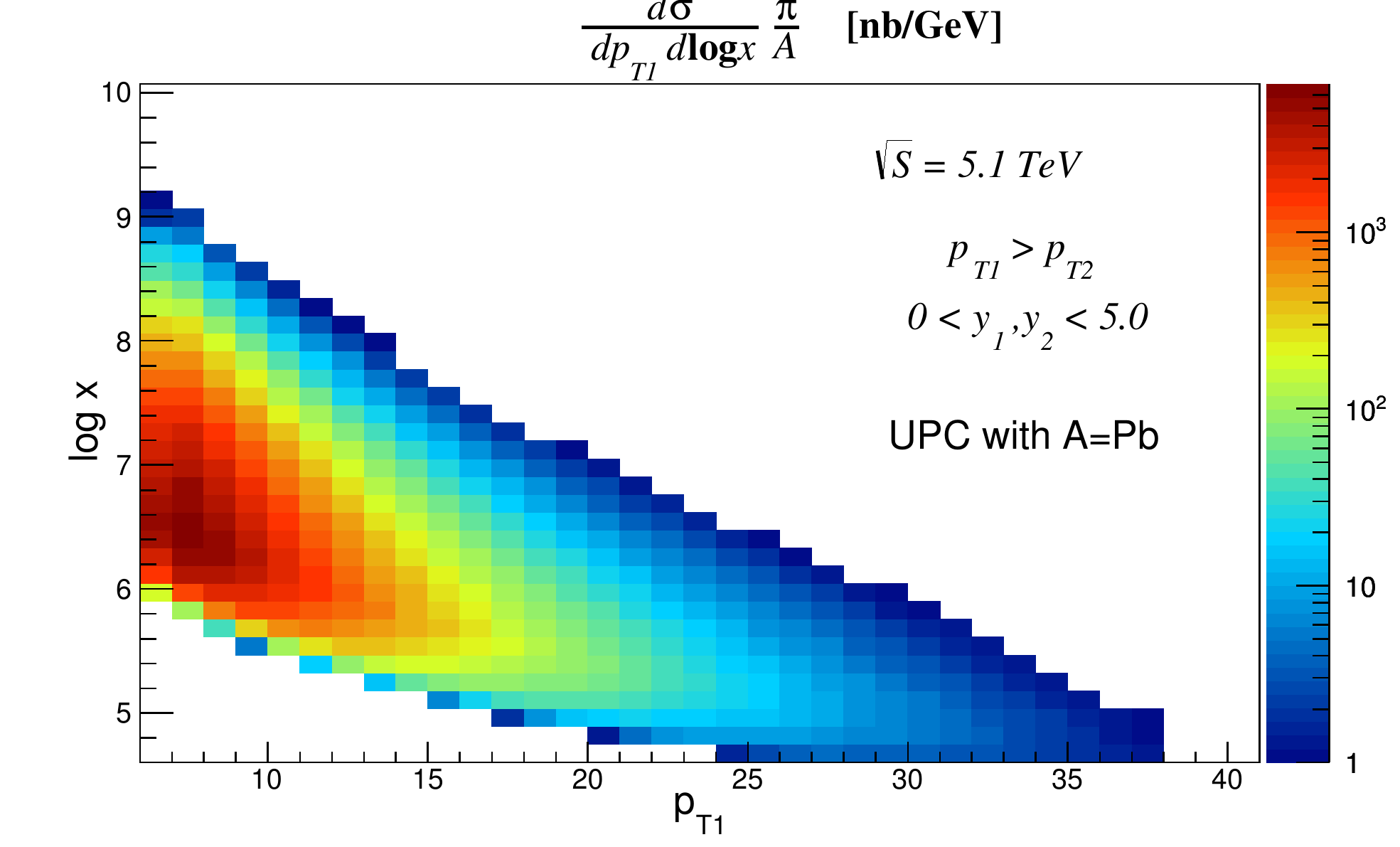}$\,\,$\includegraphics[width=6.7cm,height=6cm]{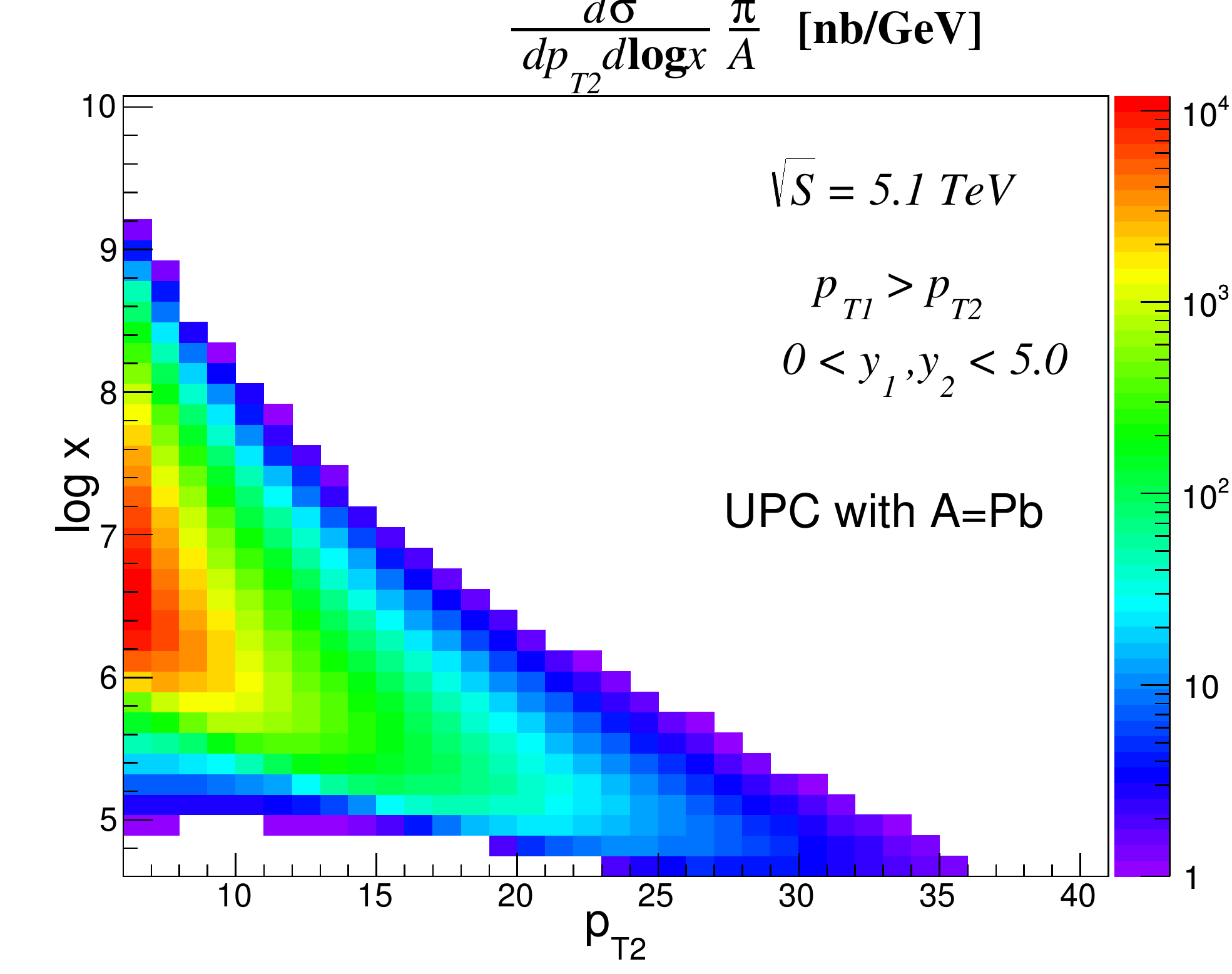}
\par\end{centering}

\caption{Phase space coverage in $p_{T}$ and the longitudinal fraction $\log x$ probed in the nucleus for leading (left)
and sub-leading (right) jets.  \label{fig:xsec-pTlogx}}
\end{figure}

In Fig.~\ref{fig:xsec-pTlogx} we gather
the information on the $p_{T}$ spectra and the probed longitudinal fractions in the nucleus $x_A$
 in one 2D plot (for leading and subleading jets) to summarize
the phase space coverage. 

\begin{figure}
\begin{centering}
\includegraphics[width=7cm]{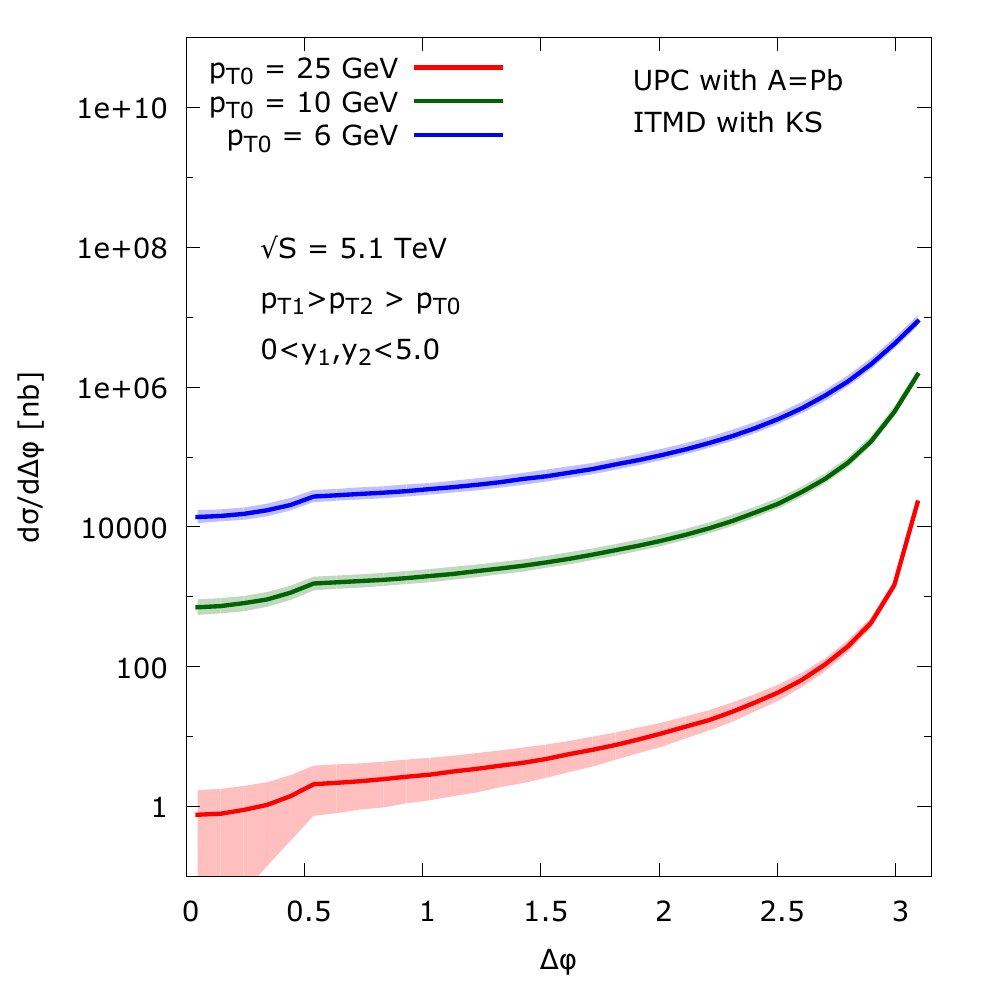}$\,\,$\includegraphics[width=7cm]{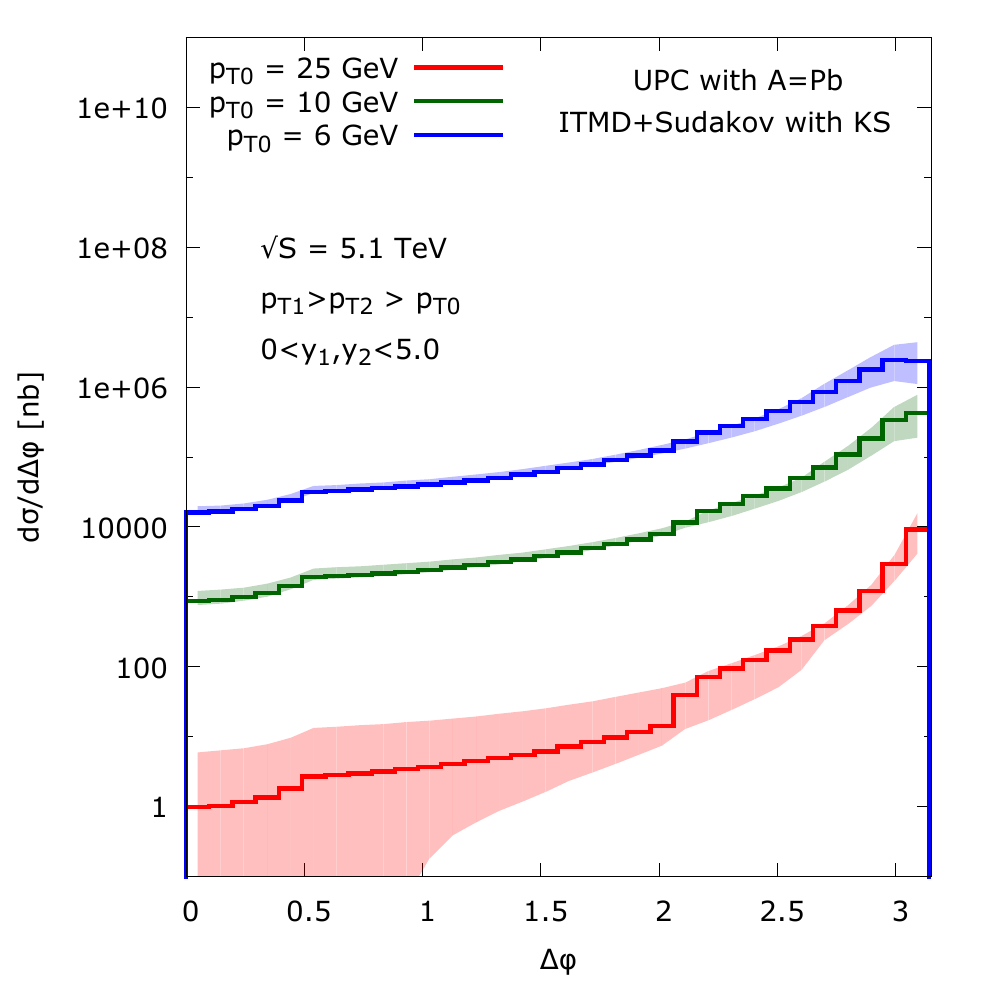}
\par\end{centering}

\caption{Differential cross sections in the azimuthal angle between the jets
with (right) and without (left) Sudakov resummation model. \label{fig:xsec-dphi}}
\end{figure}

One of the most interesting observables in the context of the small-$x$
physics are azimuthal correlations, i.e. the differential cross sections
as a function of the azimuthal angle between the two jets $\Delta\phi$.
In Fig.~\ref{fig:xsec-dphi} we calculate   this observable for various $p_{T}$
cuts. The small kinks for $\Delta\phi$ about $0.5$ are due to the
jet algorithm. Namely, when the incoming gluon has non-zero $k_T$ there is an
additional singularity, as the $k_T$ acts like an additional parton. Here, this
singularity is  regulated by the jet algorithm. The Sudakov resummation
model (right plot of Fig.~\ref{fig:xsec-dphi}) flushes the $\Delta\phi\sim\pi$
events towards the smaller values of $\Delta\phi$, as expected.

\begin{figure}
\begin{centering}
\includegraphics[width=7cm]{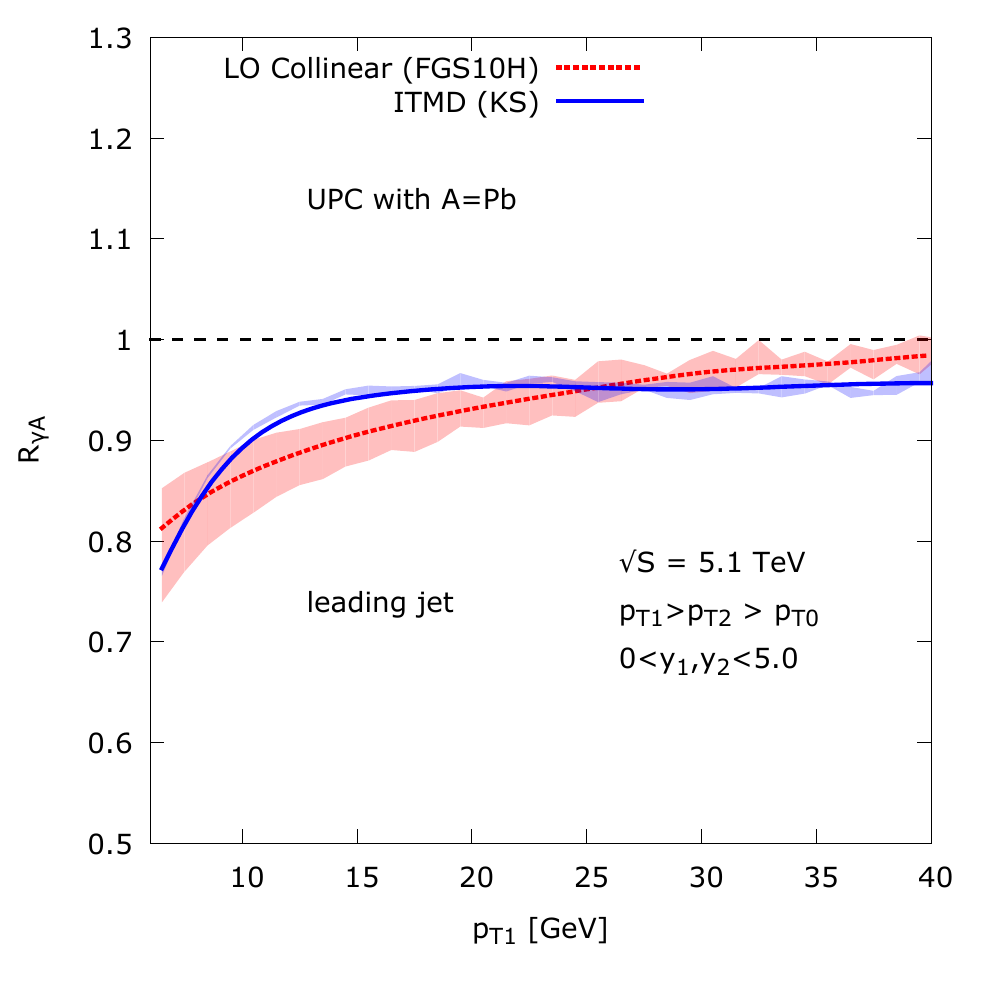}$\,\,$\includegraphics[width=7cm]{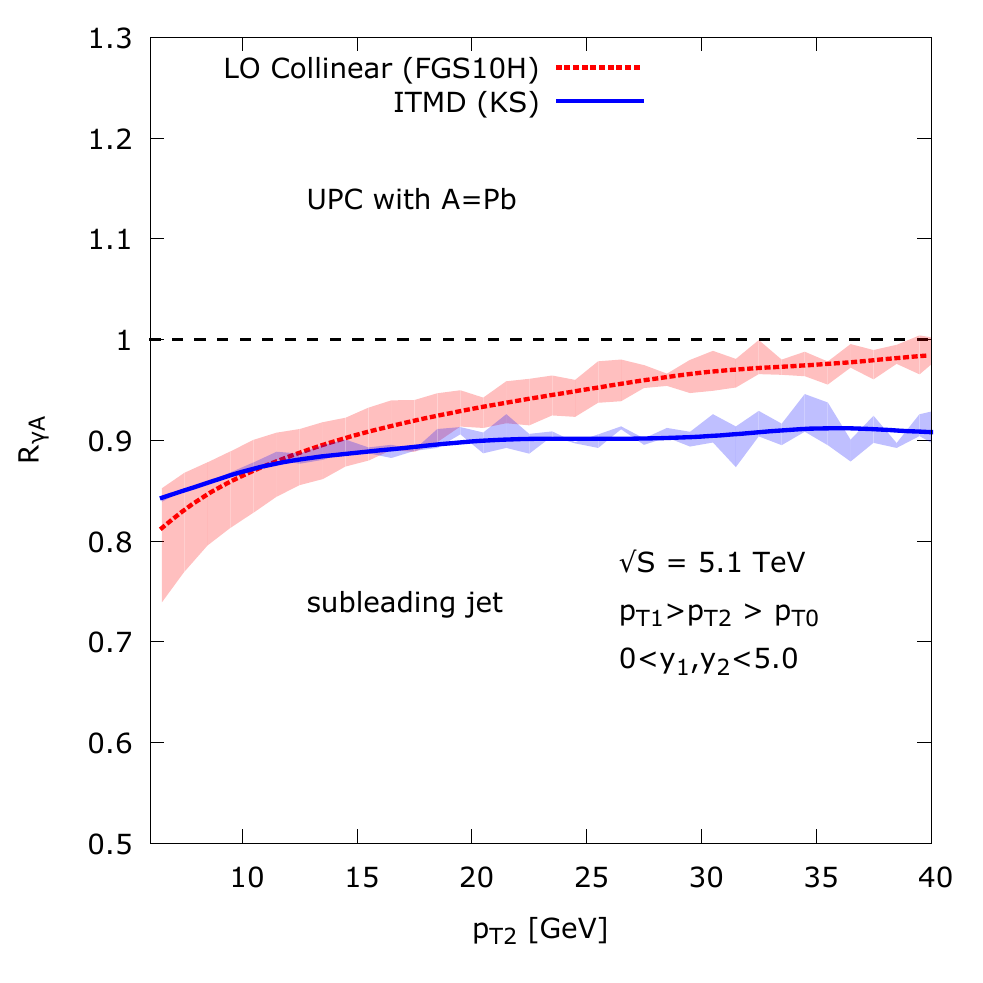}
\par\end{centering}

\begin{centering}
\includegraphics[width=7cm]{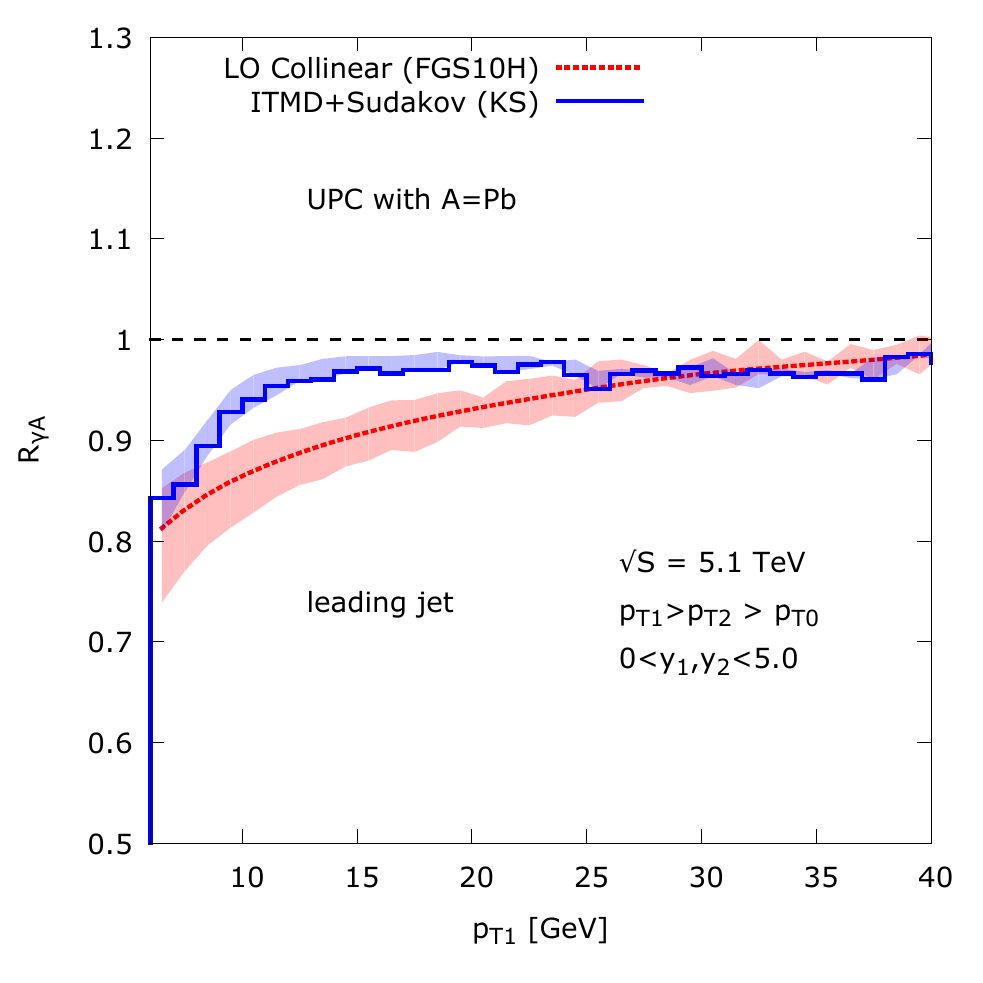}$\,\,$\includegraphics[width=7cm]{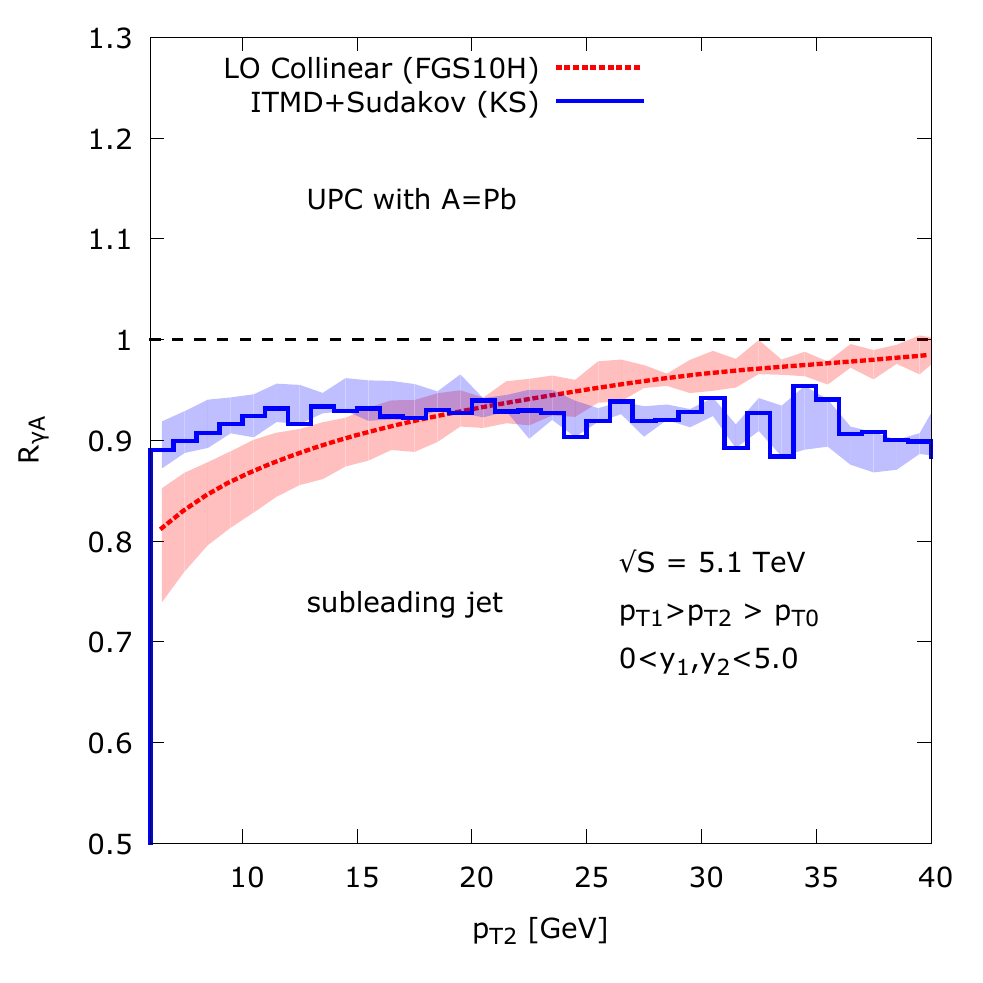}
\par\end{centering}

\caption{Nuclear modification ratio defined in Eq.~(\ref{eq:RgA}) as a function
of the transverse momentum spectra for leading (left column) and sub-leading
(right column) jets. The bottom row show the effect of the Sudakov
resummation model applied to the generated events. For comparison
we show the results from the LO collinear factorization using nuclear
PDFs with 'leading twist' shadowing. \label{fig:RgA-pT}}
\end{figure}

\begin{figure}
\begin{centering}
\includegraphics[width=7cm,height=8cm]{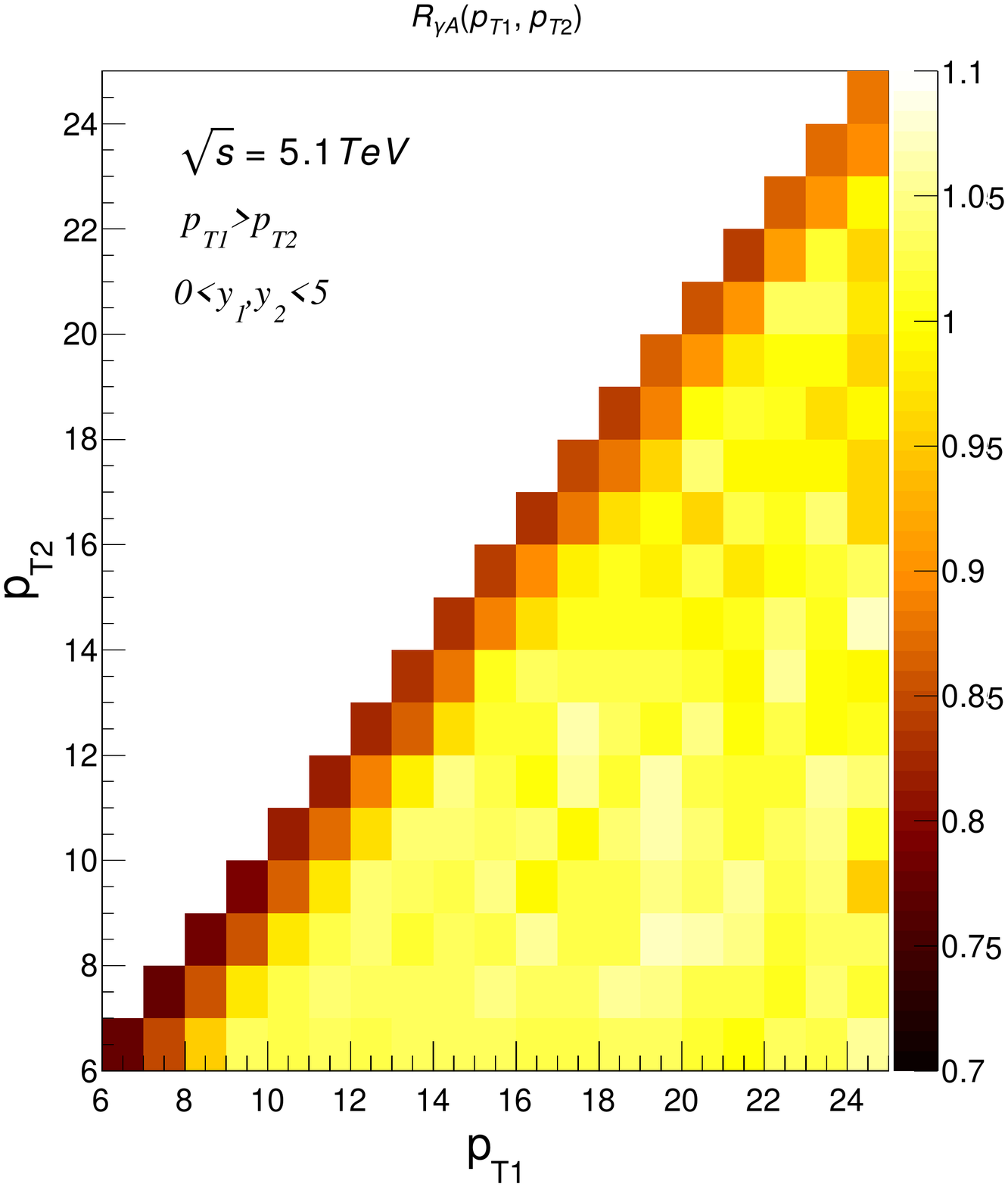}$\,\,$\includegraphics[width=7cm]{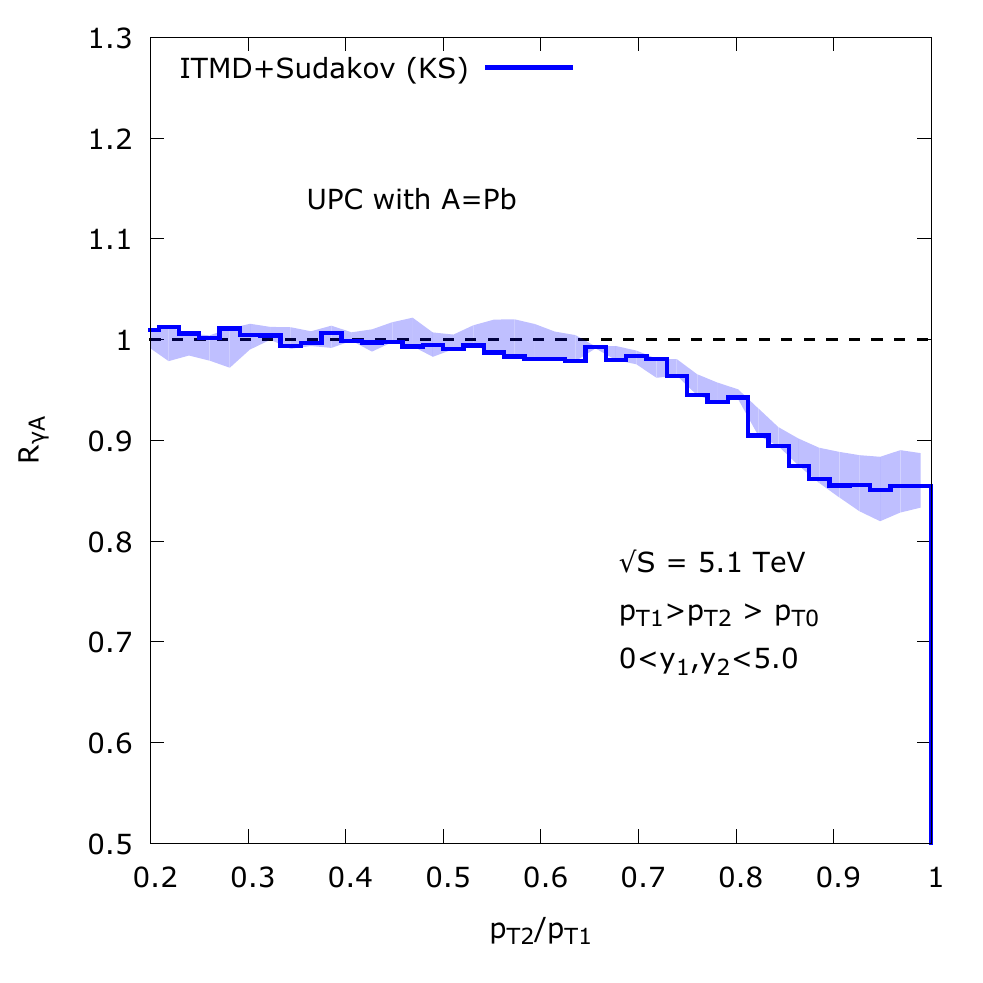}
\par\end{centering}

\caption{Nuclear modification ratio defined in Eq.~(\ref{eq:RgA}) in the
$p_{T1}-p_{T2}$ plane (left) and as a function of the $p_{T1}/p_{T2}$
ratio (right).  \label{fig:RgA-pT1pT2}}
\end{figure}

Let us now switch to discussion of the nuclear effects. We shall
study the nuclear modification ratios defined as
\begin{equation}
R_{\gamma A}=\frac{d\sigma_{AA}^{UPC}}{Ad\sigma_{Ap}^{UPC}}\,,\label{eq:RgA}
\end{equation}
that is, the photon flux in both numerator and denominator originates
from a nucleus. Let us start with $R_{\gamma A}$ as a function of
the jet $p_{T}$ (Fig.~\ref{fig:RgA-pT}) for the lowest $p_{T}$
cut studied. The maximal suppression is about $20\%$ and it slowly
 decreases with increasing $p_{T}$. The suppression of around $5-10\%$
is present through wide range of $p_{T}$, especially for the sub-leading
jet. We have compared the saturation model calculation to the leading
twist shadowing model and observe very similar suppression, which
however vanishes much faster with increasing $p_{T}$. This is most pronounced
for the sub-leading jet. The suppression factor is also shown in the
2D plot on the $p_{T1}$-$p_{T2}$ plane and as a function of the ratio
$p_{T1}/p_{T2}$ (Fig.~\ref{fig:RgA-pT1pT2}). When the Sudakov
resummation is applied the spectra change slightly. The suppression
of $R_{\gamma A}$ becomes a little bit smaller. As a result the nuclear ratio approaches the unity faster for the
leading jet spectrum. For the subleading jet the suppression also
slightly decreases, but it seems to increase for larger $p_{T}$,
although the calculation has large fluctuations there 
(the calculation is done close to the edge of the UGD grids here, so the real uncertainties are much bigger than presented).
 We note that all $R_{\gamma A}$ functions in $p_T$, both for leading and
subleading jet should approach unity for large $p_T$ (see also the discussion of
$x$ dependence at the end of this section). 

\begin{figure}
\begin{centering}
\includegraphics[width=7cm]{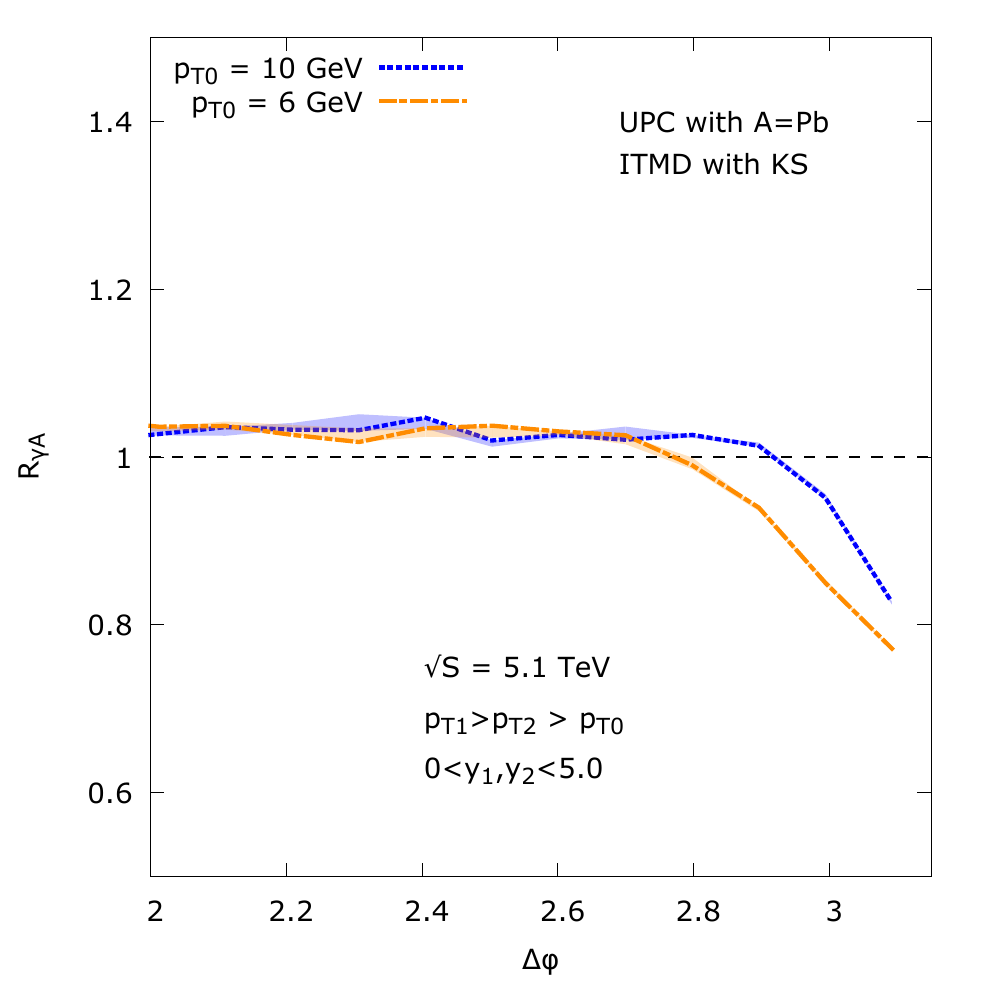}$\,\,$\includegraphics[width=7cm]{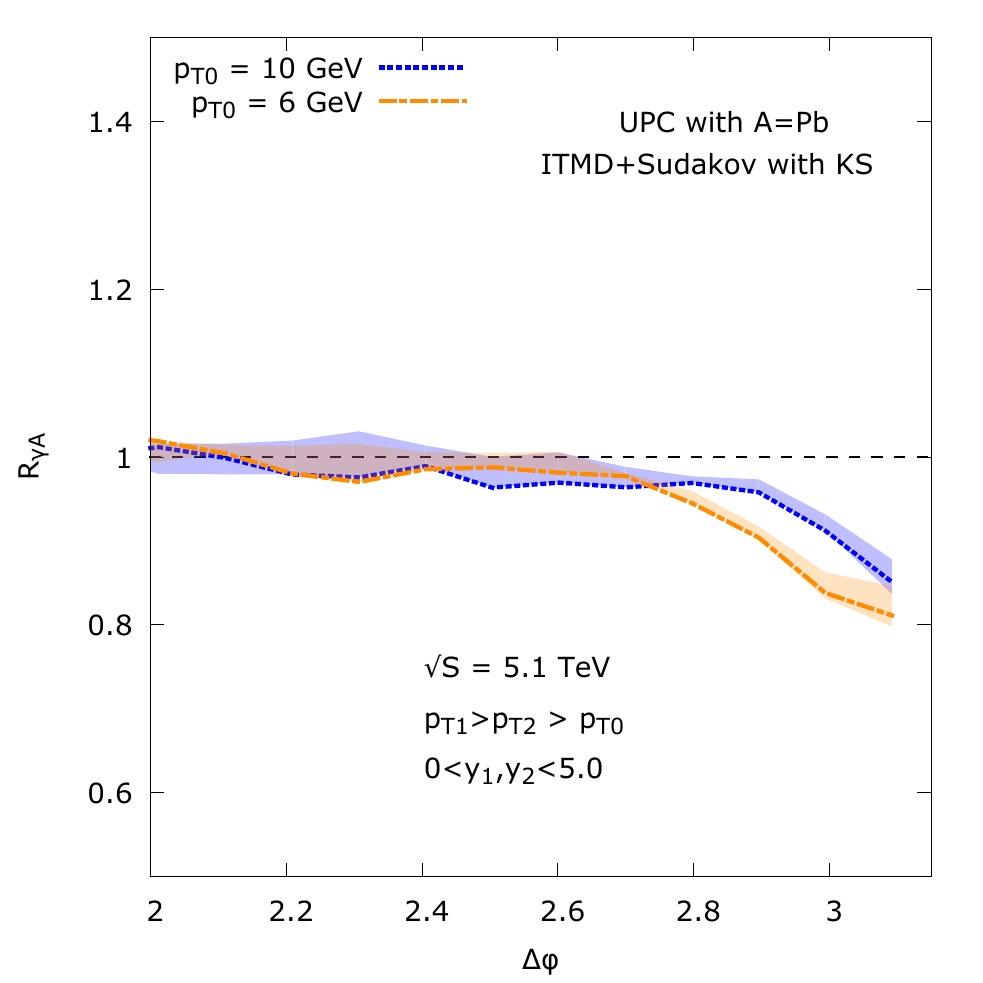}
\par\end{centering}

\caption{Nuclear modification ratio defined in Eq.~(\ref{eq:RgA}) as a function
of the azimuthal angle between the jets with (right) and without (left)
Sudakov resummation model. \label{fig:RgA-dphi}}
\end{figure}

In Fig.~\ref{fig:RgA-dphi} we present the $R_{\gamma A}$ as a function
of $\Delta\phi$ for different $p_{T}$ cuts. Again, the maximal suppression
of around $20\%$ is clearly visible for $6\,\mathrm{GeV}$ $p_{T}$
cut. For the most realistic $p_{T}$ cut of $25\,\mathrm{GeV}$ which
is not shown in the plot, the suppression was around $10\%$, but
the curve had a non-monotonic shape which was due to the grid effects,
as for that $p_{T}$ cut the $xG_{1}$ UGD is probed at relatively
large, close to the edge of the fitted parameter space.
The saturation effects are most visible close to back-to-back region
(i.e. the leading twist region) and quickly vanish with decrease
of $\Delta\phi$. The spectra show a few percent of enhancement below
 $\Delta\phi\sim 2.8$, but this is a numerical effect hidden in the UGD grids.
The situation slightly changes when the Sudakov resumation model is
used, as seen in the bottom row of the Fig.~\ref{fig:RgA-dphi}.
The suppression is spread over a slightly larger region of $\Delta\phi$
and the artificial enhancement is gone.

\begin{figure}
\begin{centering}
\includegraphics[width=7cm]{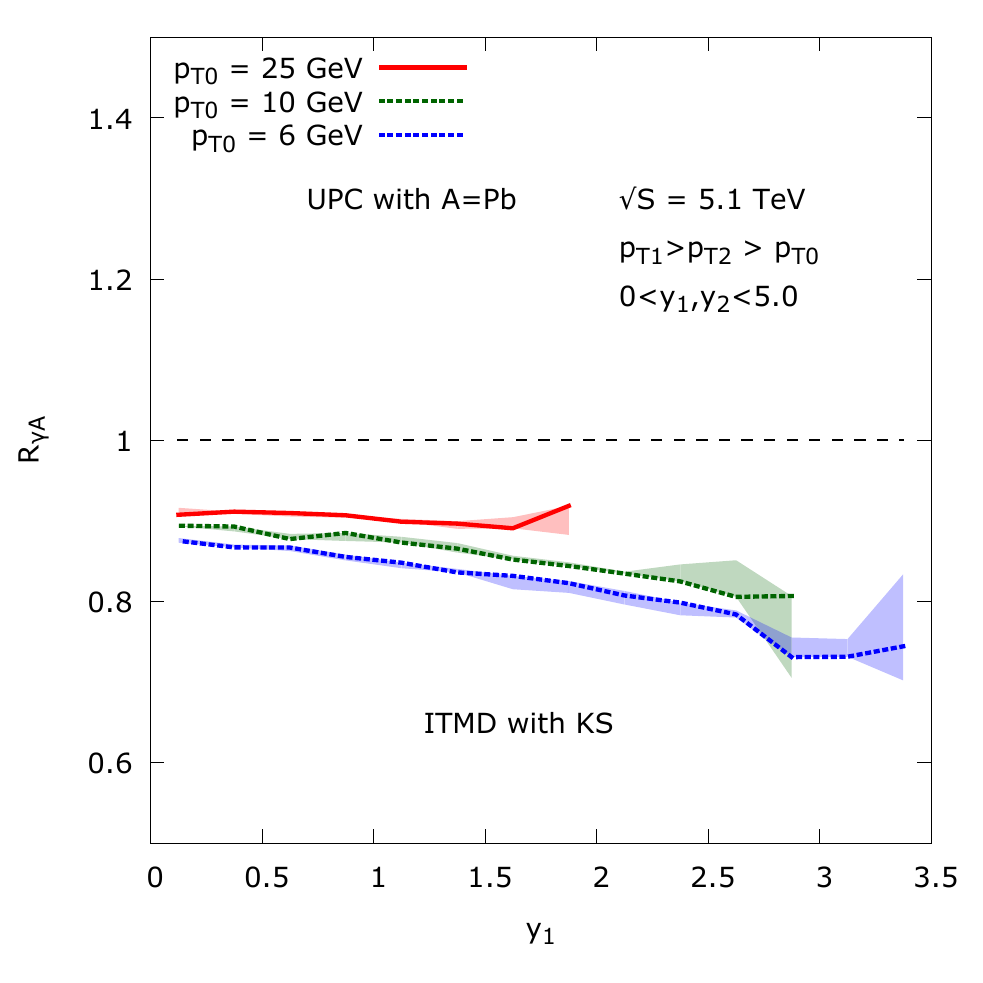}$\,\,$\includegraphics[width=7cm]{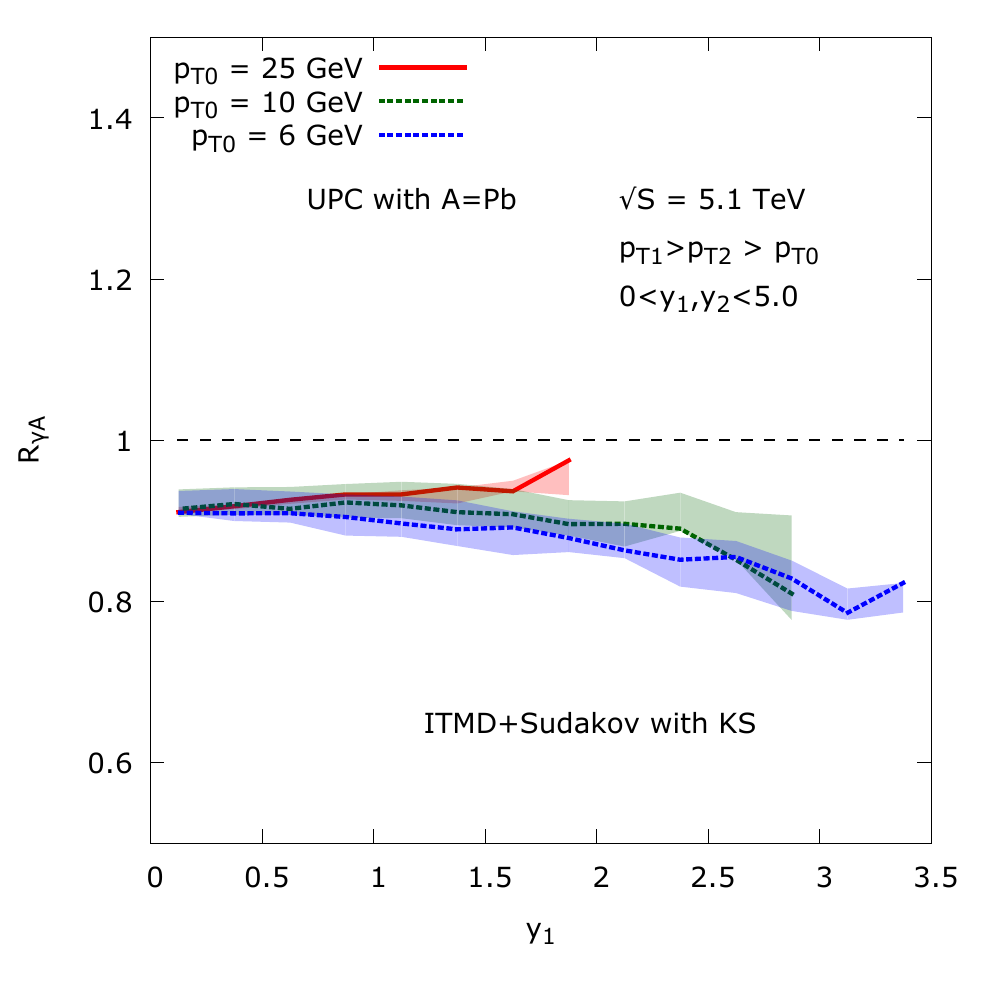}
\par\end{centering}

\begin{centering}
\includegraphics[width=12cm]{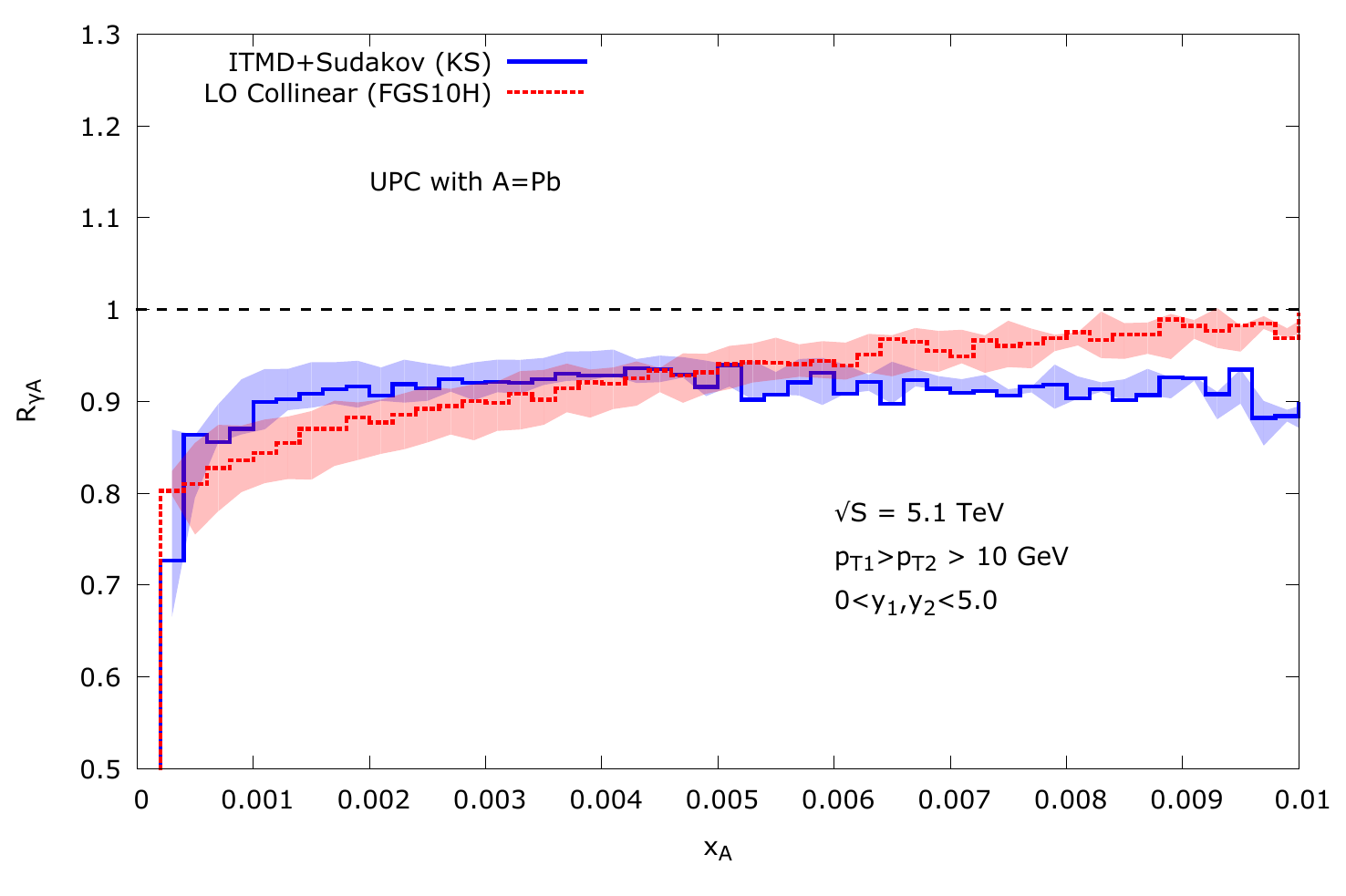}
\par\end{centering}

\caption{Nuclear modification ratio defined in Eq.~(\ref{eq:RgA}) as a function
of the rapidity of the jet without (left) and with (right) the Sudakov
resummation model. The bottom plot shows the nuclear modification ratio as a function of the longitudinal fraction $x_A$ probed in the nucleus for the $p_T$ cut $10\,\mathrm{GeV}$. \label{fig:RgA-rap}}
\end{figure}

Finally, for completeness, in Fig.~\ref{fig:RgA-rap} we show the
suppression as a function of rapidity 
(these spectra are the same for leading and subleading jets).
 The curves slowly fall off
with increase of the jet rapidity, as one could expect. After applying
the Sudakov resummation, the spectra almost do not change, but the
error bands become significantly bigger. For the $25\,\mathrm{GeV}$
$p_{T}$ cut the spectrum rises, but, again, this is the region that
involves quite large $x$, for which the UGD grids are not trustworthy.
It is interesting to note, that close to central jet production, i.e. at rather large $x_A$, there is
an initial suppression of around 10\% (we note however that there is a finite bin width of $0.25$ unit, so this statement should be taken in the average sense).
This is also clearly visible when we plot
the nuclear modification ratio as a function of the longitudinal fraction $x_A$
probed in the nucleus (Fig.~\ref{fig:RgA-rap} bottom). For definiteness, we plot
the result for the $p_T$ cut of $10\,\mathrm{GeV}$. We compare the tendency of
the saturation formalism used in the present work with the leading twist
shadowing. The calculation with saturation gives a suppression about 10\% over
the wide range of $x$: from $10^{-3}$ up to $10^{-2}$. For larger $x$ (not
shown) there are large fluctuations as we approach the edge of the phase space,
but the ratio seems to be closer to unity. For the leading twist shadowing the
ratio approaches the unity much faster (around $10^{-2}$).


\section{Summary}

\label{sec:Summary}

In this work we have investigated  potential saturation effects
in dijet production in ultra-peripheral heavy ion collisions at the LHC,
for the $5.1\,\mathrm{TeV}$ CM energy per nucleon. The quasi-real photons
are unique probes of the nucleus, as within the saturation formalism  the
Weizs\"acker-Williams~(WW) unintegrated gluon distribution is directly involved
in the dijet production process. 
The WW distribution
 has an interpretation of the gluon number density,
unlike  other similar quantities that appear at small $x$. 

In our work 
we used the ITMD formalism similar to the one used in \citep{Kotko:2015ura,VanHameren2016} for $pA$ scattering. For sufficiently
large jet $p_{T}$ it interpolates between the leading power limit of
CGC formulas and the high energy $k_{T}$-factorization. The former is
the back-to-back region of dijet imbalance and dominates the cross
section. This is also the region where the saturation effects are
strong. The latter is the region of very large imbalance, where the
linear effects are dominant. 
 There are number of advantages to this formulation:
\begin{itemize}
\item The formalism has a form of 
$k_{T}$-factorization which involves a convolution of unintegrated gluon distribution and off-shell matrix element. On phenomenology side, the usage of unintegrated gluon distributions is more convenient than
using correlators of Wilson lines. Gluon distributions can be more easily supplemented with additional effects. 
\item It involves full momentum conservation for produced final states, taking
into account the transverse momentum of the incoming gluons. This allows for a construction of Monte Carlo generators based on the formalism. 
\item Formalism is simple  compared to the full CGC
calculation, yet catching  its essential features. When using the McLerran-Venugopalan model to obtain the WW gluon distribution, the present formulation should give identical results to CGC for $\Delta\phi\sim 0$ and $\Delta\phi\sim \pi$ for large $p_T$ jets. They could differ in the intermediate region, but taking into account  general properties of $\Delta\phi$ distributions they cannot differ too much. The $p_T$ spectra should also be similar for large $p_T$.
\end{itemize}

In numerical computations we have used the unintegrated
gluon distributions which evolve  according to the nonlinear equation with subleading BFKL effects like energy conservation, running $\alpha_s$ and DGLAP correction. They were fitted to the inclusive proton HERA
data (for nucleus the nonlinear term was scaled according to the Woods-Saxon
formula) \citep{Kutak:2012rf}. By definition these are the dipole unintegrated gluon distributions.
The Weizs\"acker-Williams gluons were obtained using the Gaussian approximation
known in CGC following the procedure described in \citep{VanHameren2016}. 

The results can be summarized as follows. The suppression due to the
saturation effects is around $20\%$ at most, for the smallest $p_{T}$
cutoff of the dijet transverse momenta. This is because the probed
longitudinal fractions $x$ are not very small. In addition, for the
$p_{T}$ spectra, the saturation effects and the leading twist shadowing
look very similar. Thus, in principle it would be very difficult to
distinguish both mechanisms. 
 The main difference between the two is how fast the nuclear effects vanish
 with increase of $x$ (see Fig.~\ref{fig:RgA-rap} bottom). For the leading-twist nuclear
 shadowing this happens around $10^{-2}$, while for the saturation formalism with the WW gluon distribution
 used here it is a bigger value, very close to the edge of the phase space, so that we were unable to determine the exact value.
 The question whether one should combine both mechanisms (and whether this is possible) remains open. 
In general, the predicted nuclear
effects -- no matter of the source -- seem to be big enough to be
seen in the data. Finally we note, that discussed effect strongly depends on the centrality of the $\gamma A$ collisions. So the study of the disbalance of jets as a function of centrality appears to  be a promising strategy for exploring  the effects discussed in this paper.

\section*{Acknowledgements}

The work was supported by the Department of Energy Grants No. DE-SC-0002145,
DE-FG02-93ER40771 and by the National Science Center, Poland, Grant No. 2015/17/B/ST2/01838. P.K., K.K. and S.S. are thankful to A.~van~Hameren, C.~Marquet, E.~Petreska for fruitful discussions.
K.K. thanks the support by Narodowe Centrum Nauki with Sonata Bis grant
DEC-2013/10/E/ST2/00656.

\bibliographystyle{JHEP}
\bibliography{library}

\end{document}